\documentclass[12pt]{article}
\usepackage[centertags]{amsmath}
\usepackage{amsfonts}
\usepackage{amssymb}
\usepackage{amsthm}
\usepackage{amsmath}
\usepackage{newlfont}
\usepackage{graphicx}
\usepackage{subfig}
\usepackage{placeins}
\usepackage{caption}
\usepackage{booktabs}
\usepackage{amsthm}
\usepackage{hyperref}
\usepackage{epsfig}
\usepackage{color}
\usepackage{latexsym}
\usepackage{graphicx}
\usepackage{booktabs}
\usepackage{array}
\usepackage{rotating}
\newfont{\bb}{msbm10 at 14pt}

\usepackage{setspace}

\usepackage[left=2cm,right=2cm,top=2.3cm,bottom=2.3cm]{geometry}
\usepackage[font=small,skip=0.1pt]{caption}
\begin{document}
	
	\begin{center}{\Large\bf{The Generalized Fractional NU Method for the Diatomic Molecules in the Deng-Fan Model}}
	\end{center}
	\begin{center}
			\def\baselinestretch{1}
		M. Abu-Shady, \footnote{dr.abushady@gmail.com } T. A. Abdel-Karim, \footnote{tabdelkarim63@yahoo.com}
			and  E. M. Khokha \footnote{emadmohamed2015@gmail.com} 
			 
	\bigskip
	$^{1,2}$  Department of Mathematics and Computer Science, Faculty of Science, Menoufia University, Egypt

	$^{3}$	Department of Basic Science, Modern Academy for Engineering and Technology in Maadi, Cairo, Egypt 	
\end{center}
\begin{abstract}
A solution of the fractional ${N}$-dimensional radial Schrödinger equation (SE) with the Deng-Fan potential (DFP) is investigated by the generalized fractional NU method. The analytical formulas of energy eigenvalues and corresponding eigenfunctions for the DFP are generated. Furthermore, the current results are applied to several diatomic molecules (DMs) for the DFP as well as the shifted Deng-Fan potential (SDFP). For both the DFP and its shifted potential, the effect of the fractional parameter (${\delta}$) on the energy levels of various DMs is examined numerically and graphically. We found that the energy eigenvalues are gradually improved when the fractional parameter increases. The energy spectra of various DMs are also evaluated in three-dimensional space and higher dimensions. It's worthy to note that the energy spectrum raises as the number of dimensions increases. In addition, the dependence of the energy spectra of the DFP and its shifted potential on the reduced mass, screening parameter, equilibrium bond length, rotational and vibrational quantum numbers is illustrated. To validate our findings, we estimate the energy levels of the DFP and SDFP at the classical case (${\delta=1}$) for various DMs and found that they are entirely compatible with earlier studies.
\end{abstract}
\section{Introduction}
Fractional calculus (FC) has attracted a huge number of researchers throughout the last decades in the last and present century. The prominence of FC in diverse disciplines of science and engineering has grown considerably \cite{p9}- \cite{p11}. The foundations of FC are non-integer order differentiation and integration. In the literature, there are numerous definitions have been suggested for the fractional differential equations. The definitions of Jumarie \cite{p12}, Riemann-Liouville \cite{p13} and Caputo \cite{p14} are the most appropriate for physical circumstances and have received a lot of attention.

By using Jumarie-type derivative rules, Das et al. \cite{d1} -\cite{d2} investigated the approximate solutions of the ${N}$-dimensional fractional SE for generalized Mie-type potentials \cite{d1} and pseudoharmonic potential \cite{d2} for a typical DM, and they obtained the mass spectra of quarkonia with the Cornell potential corresponds to the fractional parameter ${\alpha=0.5}$. In addition they employed the power series approach to investigate the solution of the fractional Klein–Gordon (KG) equation with fractional scalar and vector and potentials \cite{d3}. 

Al-Raeei and El-Daher relied on the definition of Riemann–Liouville fractional derivative with a numerical technique to solve the space-dependent fractional SE for the Coulomb potential \cite{d4}, Van Der Walls potential \cite{d5}, Lennard-Jones potential \cite{d6} and Morse potential \cite{d7}. 

Recently, Khalil \cite{d8} has proposed a new definition of fractional derivative termed conformable fractional derivative (CFD) which upholds essential classical characteristics. Abdeljawad \cite{d9} extended the definition and established the fundamental notions of the CFD. Depending on the CFD, Karayer et al. \cite{d10} presented conformable fractional (CF) NU method to investigate the solutions of the SE with different potentials. 

Chung et al.\cite{d11} used the Heun function to investigate the CF SE for the Killingbeck and hyperbolic potentials. Within the context of CF quantum mechanics, the three-dimensional fractional harmonic oscillator was explored and some expectation values were obtained in Ref. \cite{d12}. The CF NU approach was used to obtain the one-dimensional KG equation for the generalized Hulthén potential \cite{d13}. In Ref. \cite{d14}, the fractional SE for a particle with position-dependent mass in an infinite potential well was studied using the CFD. In the context of the CFD, the ${N}$-dimensional radial SE was used to investigate the properties of heavy quarkonia for the dependent temperature potential \cite{d15}, Trigonometric Rosen–Morse potential \cite{d16}, hot-magnetized interaction potential \cite{d17}, and generalized Cornell potential \cite{d18}. The impact of fraction-order and dimensional number on heavy-quarkonium masses was also examined. Hammad et al. \cite{d19} used the CF NU method to develop the solutions of the CF Bohr Hamiltonian with the Kratzer potential for triaxial nuclei. Abu-Shady \cite{d20} explored the mathematical model for the Coronavirus Disease 2019 using the concept of the CFD. The fractional SE with the screened Kratzer potential \cite{d21} and Morse potential \cite{d22} was solved using the CF NU technique and the energies of several DMs were estimated for various fractional values. 

The investigation of DMs is a prominent branch of chemistry and molecular physics. Multiple authors have recently examined the solution of relativistic and nonrelativistic wave equations in order to understand the characteristics of some DMs using different molecular potentials. There is a considerable number of empirical potentials in the literature that provide a proper description of the interactions between atoms in the DM as the Kratzer \cite{p2}, Morse \cite{p3}, Pöschl-Teller \cite{p4}, Manning-Rosen \cite{p5}, Hulthén \cite{p6}, Schlöberg \cite{p7}, and Tietz-Hua potentials \cite{p8}.

The DFP is generally renowned as one of the most empirical potentials for accurately describing the DM energy spectrum and electromagnetic transitions. In 1957, Deng and Fan \cite{p1} postulated the DFP to explain the ro-vibrational spectrum of the DM, also called the general Morse potential. The DFP is regarded an empirical potential because it has the proper physical boundary conditions at the origin and infinity.
The DFP takes the form \cite{DF12}- \cite{DF7}
\begin{equation}\label{DFP}
V(r)=D_e\Big(1-\frac{be^{-\alpha r}}{1-e^{-\alpha r}}\Big)^2, \qquad b=e^{\alpha r_e}-1, \qquad r\in(0,\infty)
\end{equation}
The SDFP is also defined as \cite{DF9}
\begin{equation}\label{SDFP}
V(r)=D_e\Big(1-\frac{be^{-\alpha r}}{1-e^{-\alpha r}}\Big)^2-D_e,
\end{equation}
where ${D_e}$, ${r_e}$ and ${\alpha}$ are the dissociation energy, the equilibrium bond length and the screening parameter which indicates the radius of the potential, respectively.

Copious theoretical studies were reported to derive the solutions of the DFP and SDFP in both relativistic and non-relativistic regimes . For example, the approximate solutions of the DFP were investigated using the hypergeometric functions with the SE \cite{DF12} and with Klein–Gordon equation \cite{DF7}. Zhang et al. \cite{DF9} solved the SE with the DFP using the supersymmetric shape invariance formalism and also computed the rotational transition frequencies for HF molecule. 
By using the exact quantization rule method, Falaye et al. \cite{DF16} obtained the solutions of the SE with the DFP and calculated the ro-vibrational energy of some DMs. Ikhdair \cite{DF8} studied the solutions of the Dirac equation for the DFP with spin and pseudospin symmetries by the NU method. Oluwadare et al. \cite{DF6} solved the KG and Dirac equations using the NU method with the DFP. Mustafa \cite{DF17} calculated the vibrational energies of Li${_2}$ DM with the SE for the DFP. The NU technique was used to solve the SE with the DFP \cite{NU3} and SDFP \cite{DF2}, as well as the energy levels for a set of DMs were also reported. 

The solutions of the SE with the SDFP were found and the ro-vibrational energies for various DMs are also computed using the asymptotic iteration method \cite{DF1}, and the generalized pseudospectral method \cite{DF10}. Oluwadare and Oyewumi \cite{DF3} used the proper quantization rule to explore the approximate solutions of the SE with the SDFP, as well as the energy levels and the expectation values of DMs were investigated. The Feynman path integral method \cite{DF13} was employed to explore the approximate solutions of the SDFP. Furthermore, the energy eigenvalues and the eigenfunctions were investigated for some DMs. Omugbe \cite{DF5} used the WKB method to estimate the energy spectrum of the SE with the DFP for HCl, LiH, and ScH molecules. Nath and Roy \cite{DF11} utilized the NU method to evaluate the ro-vibrational energy and thermodynamic characteristics of the DFP for H${_2}$, LiH, HCl, and CO molecules. The numerical values of energy and thermodynamic quantities of K${_2}$ molecule with the DFP were reported by Onyenegecha et al. \cite{DF14} using the formula technique. The KG equation with the DFP was examined by the SWKB and WKB methods \cite{DF15}. Furthermore, the vibrational energies and scattering phase shift of HCl and LiH molecules were discusseed in the higher dimensions using the NU functional analysis \cite{DF4}.

The study of diverse problems of quantum mechanics in the higher dimensional space has recently piqued the interest of theoretical physicists. Since the higher dimension studies aid in a broad approach to the problem, allowing the desired outcomes in the lower dimensions to be acquired directly by dialing appropriate with ${N}$. Therefore, a wide number of works \cite{d23}-\cite{d34} have been reported to derive the energy spectra of SE solutions in ${N}$-dimensional space. 

Abu-Shady and Kaabar \cite{GFD} recently proposed a new definition for the fractional derivative known as the generalized fractional derivative (GFD). The GFD definition is regarded as a comprehensive type for the fractional derivative because it provides more features than the other definitions \cite{p12,p13,p14,d8}. Where the CFD can be simply obtained as a special case from the GFD. 

As a novel investigation, we employ the perception of the GFD to derive the bound state solutions of the ${N}$-dimensional SE with the DFP via the GF NU method motivated by the recent work \cite{GFD}. Moreover, we explore the impact of the fractional order and the dimensional number on the ro-vibrational energy spectra of multiple DMs.

This investigation is organized as follows: In Sec. 2, the theoretical tools required for this investigation are displayed. In Sec. 3, we get the bound state solutions of the ${N}$-dimensional SE for the DFP via the GFD NU method. In Sec. 4, the numerical estimates of the ro-vibrational energy levels of several DMs are reported. Finally, in Sec.5, we provide the conclusion of our investigation.

\section{Mathematical Tools}
\subsection{Overview of the fractional derivative definitions}
Here, we introduce some of the basic definitions of the fractional derivative. For ${\delta \in(n-1,n)}$, the Riemann–Liouville definition is defined as \cite{RFD}- \cite{CFD}
\begin{equation}\label{c7}
\mathcal{D}^{RL}{f(r)}=\frac{1}{\Gamma(n-\delta)}\frac{d^n}{dr^n}\int_{r_0}^r \frac{f(s)}{(x-s)^{\delta-n+1}} \, ds.
\end{equation}
and the Caputo definition is also defined as \cite{RFD}- \cite{CFD}
\begin{equation}\label{c8}
\mathcal{D}^C{f(r)}=\frac{1}{\Gamma(n-\delta)}\int_{r_0}^r \frac{f^n(s)}{(r-s)^{\delta-n+1}} \, ds.
\end{equation}
Due to the fact that the Riemann–Liouville and Caputo definitions do not satisfy some classical features such as chain, product, and quotient rules, Khalil et al suggested a new fractional derivative definition Known as the conformable fractional derivative (CFD) to overcome this problem. The CFD has the following form \cite{d8}
\begin{equation}\label{c9}
\mathcal{D}^{CF}{f(r)}=\lim_{\varphi \to 0} \frac{f(r+\varphi r^{1-\delta})-f(r)}{\varphi}; \qquad 0<\varphi \leqslant1 , \qquad\mathcal{D}^{CF}{f(0)}=\lim_{r \to 0^+} f^{CF}(r).
\end{equation}
where ${\mathcal{D}^{CF}}$ is the local fractional derivative operator which  satisfies the interesting properties that traditional fractional derivatives
do not, such as the formula of the derivative of the product or quotient of two functions and the chain rule.

More recently, Abu-Shady and Kaabar \cite{GFD} proposed a new definition for the fractional derivative known as the GFD that has the form: 
\begin{equation}\label{c10}
\mathcal{D}^{GF}{f(r)}=\lim_{\varphi \to 0} \frac{f\Big(r+\frac{\Gamma(\gamma)}{\Gamma(\gamma-\delta+1)}\varphi r^{1-\delta}\Big)-f(r)}{\varphi}; \qquad \gamma >-1, \gamma \in R^+,\qquad \mathcal{D}^{GF}{f(0)}=\lim_{r \to 0^+} f^{GF}(r).
\end{equation}
The main advantage of the GFD is that it fulfills the property: ${\mathcal{D}^\delta{\mathcal{D}^\gamma f(r)}=\mathcal{D}^{\delta+\gamma}f(r)}$, for a differentiable function ${f(r)}$, which is not satisfied by the CFD.
\subsection{Generalized fractional NU method }
In this subsection, our aim is to generalize the NU method into the framework of GFD.
The NU method \cite{NU} is utilized to solve the second-order differential equation which has the following form:
\begin{equation}\label{m1}
\mathcal{H}''(\rho)+\frac{\tilde{\tau }(\rho)}{\sigma(\rho)}\mathcal{H}' (\rho)+\frac{\tilde{\sigma}(\rho)}{\sigma^2(\rho)}\mathcal{H} (\rho),
\end{equation}
where ${\sigma(\rho)}$ and ${\tilde{\sigma}(\rho)}$ are functions of maximum second degree and ${\tilde{\tau}(\rho)}$ is a function of maximum first degree.

Now we consider the generalized fractional differential equation of the following form:
\begin{equation}\label{m2}
\mathcal{D}^\delta[\mathcal{D}^\delta\mathcal{H}(\rho)]+\frac{\tilde{\tau }(\rho)}{\sigma(\rho)}\mathcal{D}^\delta\mathcal{H} (\rho)+\frac{\tilde{\sigma}(\rho)}{\sigma^2(\rho)}\mathcal{H} (\rho),
\end{equation}
Using the key property of the GFD,
\begin{equation}\label{m20}
\mathcal{D}^\delta\mathcal{H} (\rho)=Q\rho^{1-\delta}\mathcal{H}'(\rho),
\end{equation}
\begin{equation}\label{m21}
\mathcal{D}^\delta[\mathcal{D}^\delta\mathcal{H}(\rho)]=Q^2\Big[\rho^{2(1-\delta)}\mathcal{H}''(\rho)+(1-\delta)\rho^{1-2\delta}\mathcal{H}'(\rho)\Big],
\end{equation}
where
\begin{equation}\label{m22}
Q=\frac{\Gamma(\gamma)}{\Gamma(\gamma-\delta+1)}.
\end{equation}
and substituting Eqs. (\ref{m20}) and (\ref{m21}) into Eq. (\ref{m2}) we obtain 
\begin{equation}\label{m23}
\mathcal{H}''(\rho)+\frac{Q(1-\delta)\rho^{-\delta}\sigma(\rho)+\tilde{\tau}(\rho)}{Q\rho^{1-\delta}\sigma(\rho)}\mathcal{H}' (\rho)+\frac{\tilde{\sigma}(\rho)}{Q^2\rho^{2-2\delta}\sigma^2(\rho)}\mathcal{H} (\rho),
\end{equation}
Eq. (\ref{m22}) can be reduced to a hypergeometric type equation as follows:
\begin{equation}\label{m24}
\mathcal{H}''(\rho)+\frac{\tilde{\tau }_{GF}(\rho)}{\sigma_{GF}(\rho)}\mathcal{H}' (\rho)+\frac{\tilde{\sigma}(\rho)}{\sigma_{GF}^2(\rho)}\mathcal{H} (\rho),
\end{equation}
where
\begin{equation}\label{m25}
\tilde{\tau }_{GF}(\rho)=Q(1-\delta)\rho^{-\delta}\sigma(\rho)+\tilde{\tau}(\rho), \qquad \sigma_{GF}(\rho)=Q\rho^{1-\delta}\sigma(\rho).
\end{equation}
where the subscript GF stands for the generalized fractional,
and ${\sigma(\rho)}$ and ${\tilde{\sigma}(\rho)}$ are polynomials of maximum ${2\delta}$-th degree and ${\tilde{\tau}(\rho)}$ is a function at most ${\delta}$-th degree.

Setting
\begin{equation}\label{m3}
\mathcal{H} (\rho)=\chi (\rho) \mathcal{V}  (\rho).
\end{equation}
Substituting Eq. (\ref{m3})  into Eq. (\ref{m24}), we obtain
\begin{equation}\label{m4}
\sigma_{GF}(\rho) \mathcal{V}''(\rho)+\tau_{GF}(\rho) \mathcal{V}'(\rho)+\lambda (\rho) \mathcal{V}(\rho)=0.
\end{equation}
where ${\chi(\rho)}$ satisfies the following relation
\begin{equation}\label{m5}
\frac{\chi' (\rho)}{\chi (\rho)}=\frac{\pi_{GF}(\rho)}{\sigma_{GF}(\rho)}.
\end{equation}
and
\begin{equation}\label{m9}
\lambda(\rho)=k(\rho)+\pi_{GF}'(\rho).
\end{equation}
Here, ${\mathcal{V}(\rho)}=\mathcal{V}_n (\rho)$ is a hypergeometric-type function, whose polynomial solutions are given by the Rodrigues formula
\begin{equation}\label{m6}
\mathcal{V}_n (\rho)=\frac{G_n}{\omega(\rho)}\frac{d^n}{d\rho^n}[\sigma_{GF}^n (\rho)\omega (\rho)], 
\end{equation}
where ${G_n}$ is the normalization constant, and ${\omega (\rho)}$ is a weight function that defined as follows \cite{DF11}:
\begin{equation}\label{m7}
\omega(\rho)=\Big[\sigma_{GF}(\rho)\Big]^{-1}\text{exp}\Big(\int \frac{\tau_{GF} (\rho )}{\sigma_{GF} (\rho )} \, d\rho\Big)
\end{equation}
The function ${\pi_{GF}(\rho)}$ has the following form: 
\begin{equation}\label{m8}
\pi_{GF}(\rho)=\frac{\sigma_{GF}'(\rho)-\tilde{\tau}_{GF}(\rho)}{2}\pm \sqrt{\Bigg[\frac{\sigma_{GF}'(\rho)-\tilde{\tau}_{GF}(\rho)}{2}\Bigg]^2-\tilde{\sigma}(\rho)+k(\rho)\sigma_{GF}(\rho)},
\end{equation}
The value of ${k(\rho)}$ can be determined if the function under the square root is the square of a polynomial. Thus, the equation of the eigenvalues can be obtained from the following relation: 
\begin{equation}\label{m10}
\lambda(\rho)=\lambda_n(\rho)=-n\Big[\tau_{GF}'(\rho)+\frac{(n-1)}{2}\sigma_{GF}''(\rho)\Big],
\end{equation}
where 
\begin{equation}\label{m11}
\tau_{GF}(\rho)=\tilde{\tau}_{GF}(\rho)+2\pi_{GF}(\rho).
\end{equation}
Finally, the eigenfunctions ${\mathcal{H} (\rho)}$ can be determined from Eq. (\ref{m3}) using Eq. (\ref{m5}) and Eq. (\ref{m6}).

\section{Solution of the generalized fractional SE for the DFP}
The ${N}$-dimensional radial SE for a DM can be written as \cite{d23}- \cite{d34}
\begin{equation}\label{x1}
\Biggl\{\frac{d^2}{dr^2}+\frac{N-1}{r}\frac{d}{dr}-\frac{l(l+N-2)}{r^2}+\frac{2\mu}{\hbar^2}\Big(E-V(r)\Big)\Biggl\}\phi(r)=0,
\end{equation}
where ${\mu, \hbar,E}$ and ${N}$ are the reduced mass, the reduced
Planck's constant, the energy spectrum and the dimensional number respectively. Let us set, 
\begin{equation}\label{x2}
\phi(r)=\frac{1}{r^\frac{N-1}{2}} F(r).
\end{equation}
Eq. (\ref{x1}) becomes
\begin{equation}\label{x3}
\frac{d^2F(r)}{dr^2}+\Bigg[\frac{2\mu}{\hbar^2}\Big(E-V(r)\Big)-\frac{(\eta^2-\frac{1}{4})}{r^2}\Bigg]F(r)=0,
\end{equation}
with
\begin{equation}\label{x4}
\eta=l+\frac{N-2}{2},
\end{equation}
Here we consider a general form of the DFP as follows:
\begin{equation}\label{GDFP}
V(r)=D_e\Big(1-\frac{be^{-\alpha r}}{1-e^{-\alpha r}}\Big)^2+V_0, \qquad b=e^{\alpha r_e}-1, \qquad r\in(0,\infty)
\end{equation}
The DFP can be found by setting ${V_0=0}$. In the case if ${V_0=-D_e}$, Eq. (\ref{GDFP}) turns to the SDFP.
By inserting Eq. (\ref{GDFP}) into Eq. (\ref{x3}), we obtain:   
\begin{equation}\label{x5}
\frac{d^2F(r)}{dr^2}+\Biggl\{\frac{2\mu}{\hbar^2}\Bigg[E-D_e\Big(1-\frac{be^{-\alpha r}}{(1-e^{-\alpha r})}\Big)^2-V_0\Bigg]-\frac{(\eta^2-\frac{1}{4})}{r^2}\Bigg)\Biggl\}F(r)=0,
\end{equation}
Now we utilize a proper approximation for the term ${1/r^2}$ as \cite{DF11}
\begin{equation}\label{x6}
\frac{1}{r^2}\approx\alpha^2\Bigg[c_0+\frac{e^{-\alpha r}}{(1-e^{-\alpha r})}+\frac{e^{-2\alpha r}}{(1-e^{-\alpha r})^2}\Bigg]; \qquad c_0=\frac{1}{12}.
\end{equation}
Inserting Eq. (\ref{x6}) into Eq. (\ref{x5}), we obtain:
\begin{equation}\label{x7}
\frac{d^2F(r)}{dr^2}+\Biggl\{\frac{2\mu}{\hbar^2}\Bigg[E-D_e\Big(1-\frac{be^{-\alpha r}}{(1-e^{-\alpha r})}\Big)^2-V_0\Bigg]-\alpha^2\Big(\eta^2-\frac{1}{4}\Big)\Bigg[c_0+\frac{e^{-\alpha r}}{(1-e^{-\alpha r})}+\frac{e^{-2\alpha r}}{(1-e^{-\alpha r})^2}\Bigg]\Biggl\}F(r)=0,
\end{equation}
By defining the variable ${\rho=e^{-\alpha r}}$, Eq. (\ref{x7}) turns to 
\begin{equation}\label{x9}
F''(\rho)+\frac{(1-\rho)}{\rho(1-\rho)}F'(\rho)+\frac{1}{\rho^2(1-\rho)^2}\Big[-\mathcal{A}\rho^2+\mathcal{B}\rho-\mathcal{C}\Big]F(\rho)=0,
\end{equation}
where
\begin{equation}\label{x10}
\mathcal{A}=c_0\Big(\eta^2-\frac{1}{4}\Big)+\frac{2\mu}{\alpha^2\hbar^2}\Big[V_0+D_e(b+1)^2\Big]-\epsilon,
\end{equation}
\begin{equation}\label{x11}
\mathcal{B}=(2c_0-1)\Big(\eta^2-\frac{1}{4}\Big)+\frac{4\mu}{\alpha^2\hbar^2}\Big[V_0+D_e(b+1)\Big]-2\epsilon,
\end{equation}
\begin{equation}\label{x12}
\mathcal{C}=c_0\Big(\eta^2-\frac{1}{4}\Big)+\frac{2\mu}{\alpha^2\hbar^2}\Big(V_0+D_e\Big)-\epsilon,
\end{equation}
with
\begin{equation}\label{x40}
\epsilon=\frac{2\mu E}{\alpha^2\hbar^2}
\end{equation}
The generalized fractional form of the SE for the DFP can be stated by changing integer orders with fractional orders in Eq. (\ref{x9}) as follows:
\begin{equation}\label{x13}
\mathcal{D}^\delta\mathcal{D}^\delta F(\rho)+\frac{(1-\rho^\delta)}{\rho^\delta(1-\rho^\delta)}\mathcal{D}^\delta F(\rho)+\frac{1}{\rho^{2\delta}(1-\rho^\delta)^2}\Big[-\mathcal{A}\rho^{2\delta}+\mathcal{B}\rho^\delta-\mathcal{C}\Big]F(\rho)=0,
\end{equation}
Substituting Eqs. (\ref{m20}) and (\ref{m21}) into Eq. (\ref{x13}), we obtain
\begin{equation}\label{x14}
F''(\rho)+\frac{\Big[1+Q(1-\delta)\Big](1-\rho^\delta)}{Q\rho(1-\rho^\delta)}F'(\rho)+\frac{1}{Q^2\rho^2(1-\rho^\delta)^2}\Big[-\mathcal{A}\rho^{2\delta}+\mathcal{B}\rho^\delta-\mathcal{C}\Big]F(\rho)=0,
\end{equation}
By comparing Eq. (\ref{x14}) with Eq. (\ref{m24}), we get the following functions:
\begin{equation}\label{x16}
\tilde{\tau}_{GF}(\rho)=\Big(1+Q(1-\delta)\Big)(1-\rho^\delta),  \qquad   \sigma_{GF}(\rho)=Q\rho(1-\rho^\delta),  \qquad  \tilde{\sigma}_{GF}(\rho)=-\mathcal{A}\rho^{2\delta}+\mathcal{B}\rho^\delta-\mathcal{C}.
\end{equation}
The function ${\pi_{GF}(\rho)}$ can be defined after substituting Eq. (\ref{x16}) into Eq. (\ref{m8}) as follows:
\begin{equation}\label{x17}
\begin{split} 
&\pi_{GF}(\rho)=\frac{(Q\delta-1)+(1-2Q\delta)\rho^\delta}{2}\pm\\&\sqrt{\Big[\frac{(1-2Q\delta)^2}{4}+\mathcal{A}-Qk\rho^{1-\delta}\Big]\rho^{2\delta}+\Big[\frac{(Q\delta-1)(1-2Q\delta)}{2}-\mathcal{B}+Qk\rho^{1-\delta}\Big]\rho^\delta+\Big[\frac{(Q\delta-1)^2}{4}+\mathcal{C}\Big]}.
\end{split} 
\end{equation}
Eq. (\ref{x17}) can be written in the following form:
\begin{equation}\label{x18}
\pi_{GF}(\rho)=\frac{(Q\delta-1)+(1-2Q\delta)\rho^\delta}{2}\pm\sqrt{P\rho^{2\delta}+W\rho^\delta+R},
\end{equation}
where
\begin{equation}\label{x19}
P=T_1-Qk\rho^{1-\delta}, \qquad W=T_2+Qk\rho^{1-\delta}, \qquad R=T_3,
\end{equation}
with
\begin{equation}\label{x20}
T_1=\frac{(1-2Q\delta)^2}{4}+\mathcal{A}, \qquad T_2=\frac{(Q\delta-1)(1-2Q\delta)}{2}-\mathcal{B}, \qquad T_3=\frac{(Q\delta-1)^2}{4}+\mathcal{C}.
\end{equation}
We can get the two possible roots of ${k}$ by employing the condition that the discriminant of the expression under the square root of Eq. (\ref{x18}) must be zero. Hence,
\begin{equation}\label{x21}
k_\pm=\Lambda\Big[-(T_2+2T_3)\pm2\sqrt{T_3(T_1+T_2+T_3)}\Big]\rho^{\delta-1}; \qquad \Lambda=\frac{1}{Q}.
\end{equation}
Inserting Eq. (\ref{x21}) into Eq. (\ref{x18}), we have
\begin{equation}\label{x22}
\pi_{GF}(\rho)=\frac{(Q\delta-1)+(1-2Q\delta)\rho^\delta}{2}\pm\begin{cases}
	\big(\sqrt{T_3}-\sqrt{T_1+T_2+T_3}\big)\rho^\delta-\sqrt{T_3},  \qquad  k=k_+ \\
	\big(\sqrt{T_3}+\sqrt{T_1+T_2+T_3}\big)\rho^\delta-\sqrt{T_3},  \qquad  k=k_- \\
\end{cases}.
\end{equation}
We select the negative sign in Eq. (\ref{x22}) in order to obtain a physically acceptable solution. Hence, the function ${\pi_{GF}(\rho)}$ becomes 
\begin{equation}\label{x23}
\pi^-_{GF}(\rho)=\frac{(Q\delta-1)+(1-2Q\delta)\rho^\delta}{2}-\Big[\big(\sqrt{T_3}+\sqrt{T_1+T_2+T_3}\big)\rho^\delta-\sqrt{T_3}\Big],
\end{equation}
and
\begin{equation}\label{x24}
k_-=-\Lambda\Big[T_2+2T_3+2\sqrt{T_3(T_1+T_2+T_3)}\Big]\rho^{\delta-1}.
\end{equation}
Therefore, the functions ${\lambda(\rho), \tau_{GF}(\rho)}$ and ${\lambda_n(\rho)}$ can be found as follows:
\begin{equation}\label{x25}
\lambda(\rho)=\Bigg[-\Lambda(T_2+2T_3)+\delta\Bigg(\frac{1}{2}\Big(1-2Q\delta\Big)-\sqrt{T_3}\Bigg)-\sqrt{T_1+T_2+T_3}\Big(\delta+2\Lambda\sqrt{T_3}\Big)\Bigg]\rho^{\delta-1},
\end{equation}
\begin{equation}\label{x26}
\tau_{GF}(\rho)=\Big(2\sqrt{T_3}+Q\Big)-\Big[Q(\delta+1)+2\Big(\sqrt{T_3}+\sqrt{T_1+T_2+T_3}\Big)\Big]\rho^\delta,
\end{equation}
\begin{equation}\label{x27}
\lambda_n(\rho)=n\delta\Big[\frac{Q(n+1)(\delta+1)}{2}+2\Big(\sqrt{T_3}+\sqrt{T_1+T_2+T_3}\Big)\Big]\rho^{\delta-1}.
\end{equation}
By equating Eqs. (\ref{x25}) and (\ref{x27}), the energy spectrum formula for a DM in the ${N}$ dimensional space can be written as follows: 
\begin{equation}\label{x28}
E^N_{nl}=\frac{\alpha^2\hbar^2}{2\mu}\Biggl\{\xi_3-\Bigg(\frac{\delta\Big[\frac{1}{2}\Big(1-2Q\delta-Qn(n+1)(\delta+1)\Big)-(2n+1)\sqrt{\xi_1+\xi_2+\xi_3}-\Lambda(\xi_2+2\xi_3)\Big]}{\Big(\delta(2n+1)+2\Lambda\sqrt{\xi_1+\xi_2+\xi_3}\Big)}\Bigg)^2\Biggl\},
\end{equation}
where
\begin{equation}\label{x29}
\xi_1=\frac{(1-2Q\delta)^2}{4}+c_0\Big(\eta^2-\frac{1}{4}\Big)+\frac{2\mu}{\alpha^2\hbar^2}\Big[V_0+D_e(b+1)^2\Big],
\end{equation}
\begin{equation}\label{x30}
\xi_2=\frac{(Q\delta-1)(1-2Q\delta)}{2}-(2c_0-1)\Big(\eta^2-\frac{1}{4}\Big)-\frac{4\mu}{\alpha^2\hbar^2}\Big[D_e(b+1)+V_0\Big],
\end{equation}
\begin{equation}\label{x31}
\xi_2=\frac{(Q\delta-1)^2}{4}+c_0\Big(\eta^2-\frac{1}{4}\Big)+\frac{2\mu}{\alpha^2\hbar^2}\Big(D_e+V_0\Big).
\end{equation}
Now we determine the corresponding eigenfunctions. Using Eq. (\ref{m5}) the function ${\chi(\rho)}$ is 
\begin{equation}\label{x32}
\chi (\rho)=\rho^{\Big(\frac{(Q\delta-1)}{2}+\sqrt{\xi_3-\epsilon}\Big)}\Big(1-\rho^\delta\Big)^{\Big({\frac{1}{2}+\frac{\Lambda}{\delta}\sqrt{\xi_1+\xi_2+\xi_3}}\Big)}.
\end{equation}
The weight function ${\omega(\rho)}$ is found using Eq. (\ref{m7}) as
\begin{equation}\label{x33}
\omega (\rho)=\Lambda\rho^{\Big(2\Lambda\sqrt{\xi_3-\epsilon}\Big)}\Big(1-\rho^\delta\Big)^{\Big({\frac{2\Lambda}{\delta}\sqrt{\xi_1+\xi_2+\xi_3}}\Big)}.
\end{equation}
Using Eq. (\ref{m6}), the expression for the function ${\mathcal{V}_n (\rho)}$ takes the form 
\begin{equation}\label{x34}
\mathcal{V}_n (\rho)=G_n\rho^{\Big(-2\Lambda\sqrt{\xi_3-\epsilon}\Big)}\Big(1-\rho^\delta\Big)^{\Big({\frac{-2\Lambda}{\delta}\sqrt{\xi_1+\xi_2+\xi_3}}\Big)}\frac{d^n}{d\rho^n}\Bigg[Q^n\rho^{\Big(n+2\Lambda\sqrt{\xi_3-\epsilon}\Big)}\Big(1-\rho^\delta\Big)^{\Big({n+\frac{2\Lambda}{\delta}\sqrt{\xi_1+\xi_2+\xi_3}}\Big)}\Bigg].
\end{equation}
Finally, using Eq. (\ref{m3}), we get the complete solution of Eq. (\ref{x9}) as
\begin{equation}\label{x35}
F (\rho)=G_n\rho^{\Lambda\Big(\frac{(Q\delta-1)}{2}-\sqrt{\xi_3-\epsilon}\Big)}\Big(1-\rho^\delta\Big)^{\Big({\frac{1}{2}-\frac{\Lambda}{\delta}\sqrt{\xi_1+\xi_2+\xi_3}}\Big)}\frac{d^n}{d\rho^n}\Bigg[Q^n\rho^{\Big(n+2\Lambda\sqrt{\xi_3-\epsilon}\Big)}\Big(1-\rho^\delta\Big)^{\Big({n+\frac{2\Lambda}{\delta}\sqrt{\xi_1+\xi_2+\xi_3}}\Big)}\Bigg].
\end{equation}
\begin{figure}[!h]
	\begin{center}
		\begin{minipage}{.45\textwidth}
			(a) \centering
			\includegraphics[width=1.08\linewidth]{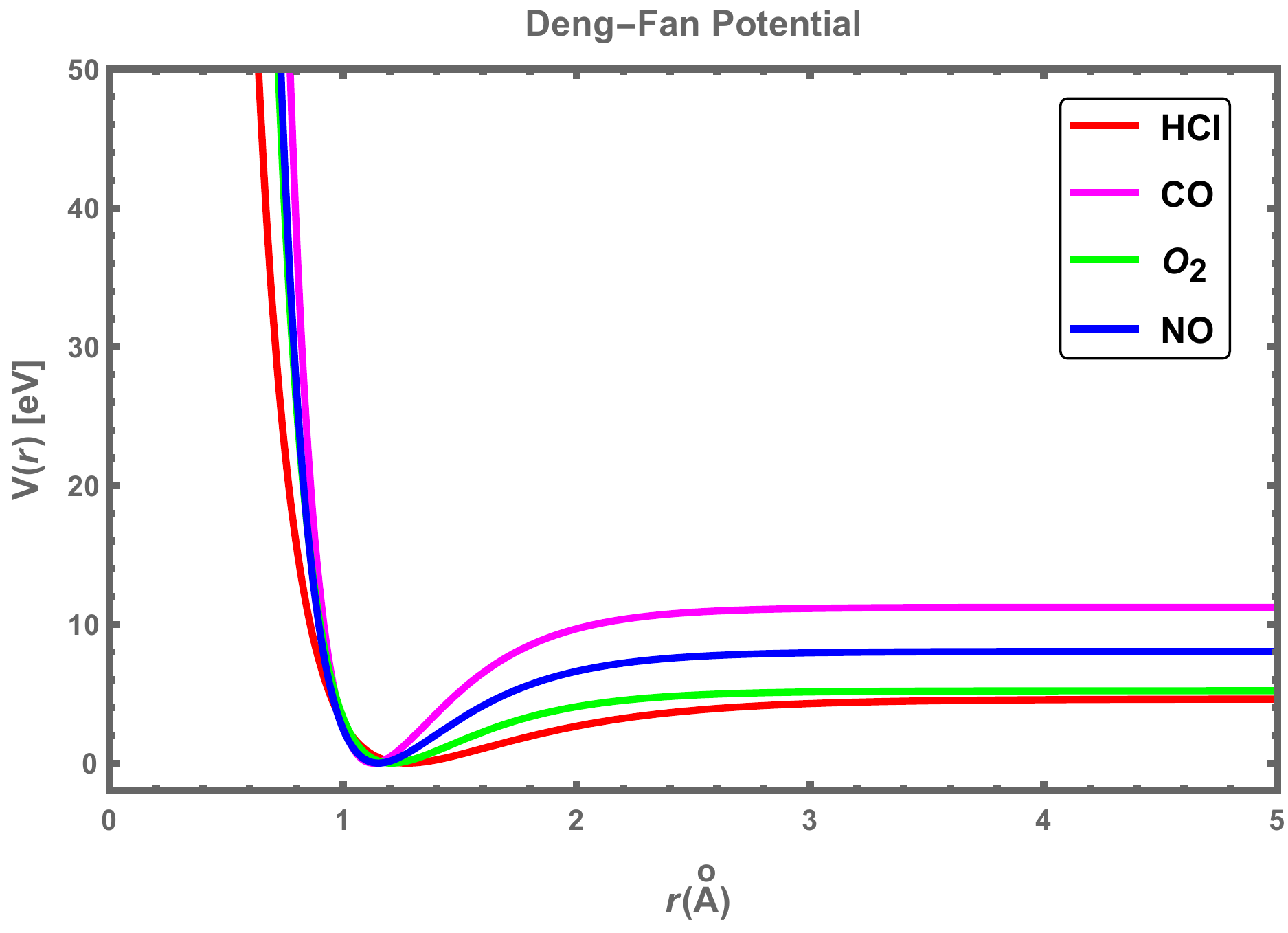}
		\end{minipage}
		\quad	
		\begin{minipage}{.45\textwidth}
			(b) \centering
			\includegraphics[width=1.08\linewidth]{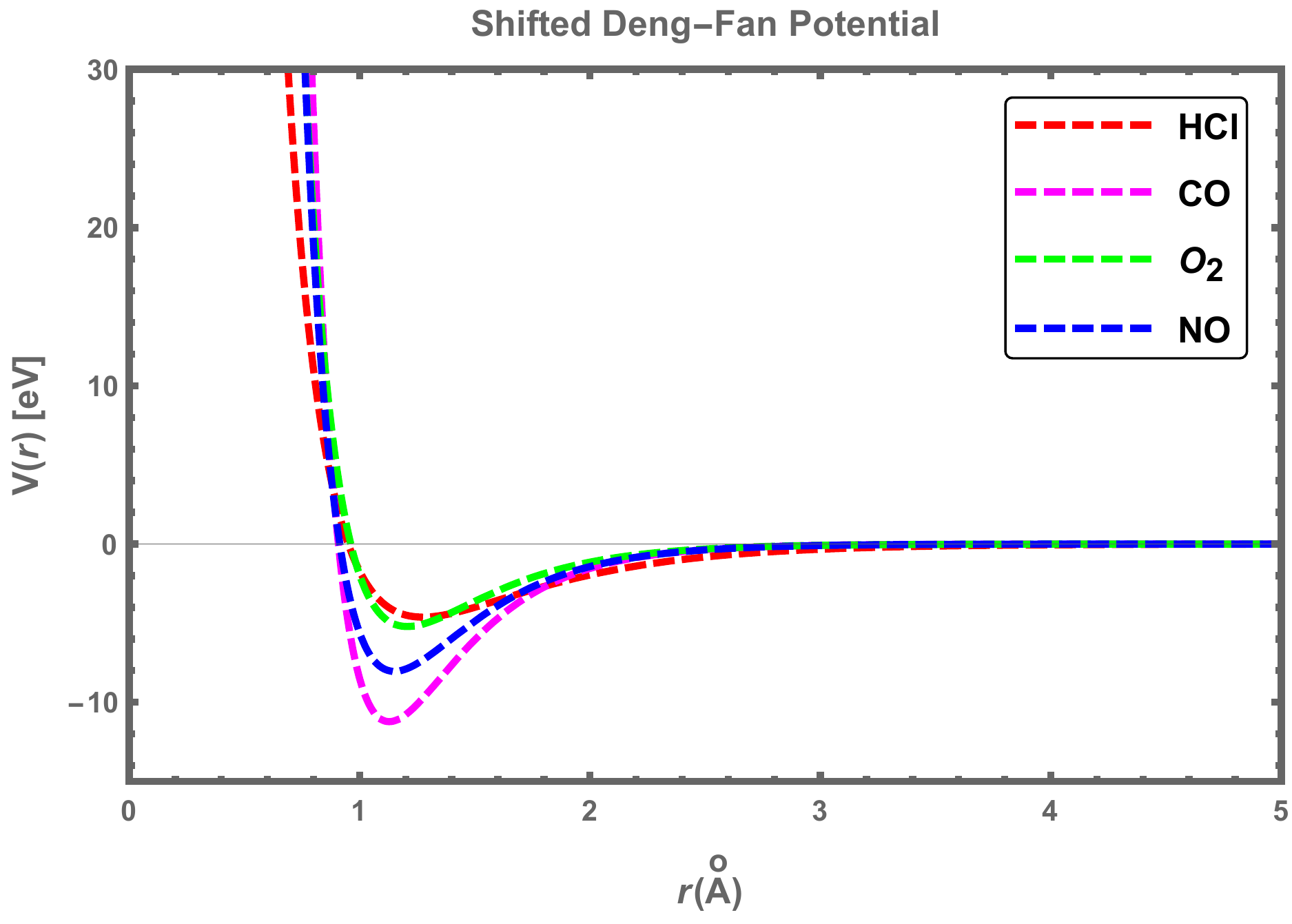}
		\end{minipage}
		\quad	
	\end{center}
	\caption{Shape of DFP and SDFP for some DMs.}\label{f1}
\end{figure}
\begin{table}[!h]
	\footnotesize
	\centering
	\caption{Spectroscopic parameters for various DMs \cite{DF3}. }\label{tab:t0}
	\begin{tabular}{ccccc}
		\hline
		Molecule & ${r_e}$ (${\overset{o}{A}}$)& ${\alpha}$ (${\overset{o}{A}^{-1}}$)& ${\mu}$ (a.m.u.) & ${D_e}$ ${(cm^{-1})}$\\
		\hline
		$ \begin{array}{c}
		\text{NO}\\
		\text{CO}\\
		 \text I_2\\
		\text N_2\\
		\text O_2\\
		\text H_2\\
		\text{HF}\\
		\text{LiH}\\
		\text{ScH}\\
		\text{HCl}\\
		\end{array}$
		&$\begin{array}{c}
		1.1508\\
		1.1282\\
		2.6620\\
		1.0940\\
		1.2070\\
		0.7416\\
		0.9170\\
		1.5956\\
		1.7080\\
		1.2746\\
		\end{array}$
		&$\begin{array}{c}
		2.7534\\
		2.2994\\
		1.8643\\
		2.6989\\
		2.6636\\
		1.9426\\
		2.2266\\
		1.1280\\
		1.5068\\
		1.8677\\
		\end{array}$
		&$\begin{array}{c}
		7.468441\\
		6.860586\\
		63.452235\\
		7.00335\\
		7.997457504\\
		0.50391\\
		0.96367\\
		0.8801221\\
		10.682771\\
		0.9801045\\
		\end{array}$
		&$\begin{array}{c}
		64877.06229\\
	    87471.42567\\
	    12758.0129\\
		96288.03528\\
		41591.26201\\
	    38267.78314\\
	    49382\\
	    20287.13295\\
	    36778.8836\\
		37255.24414\\
	    \end{array}$
		\\
		\hline
		\hline
	\end{tabular}
\end{table}
\begin{table}[!h]
	\footnotesize
	\caption{${E_{nl}}$ (eV) of DFP for CO molecule.}\label{tab:h1}
		\begin{tabular}{ccccccccc}
		\hline
		&&\multicolumn{3}{c}{${N=3}$}	 &${N=3}$&\multicolumn{3}{c}{${\delta=1}$}
		\\
		\cmidrule(lr){3-5} \cmidrule(lr){7-9}
		${n}$ &	${l}$ & ${\delta=0.2}$ & ${\delta=0.5}$ & ${\delta=0.8}$ & ${\delta=1}$ & ${N=4}$ & ${N=5}$ & ${N=6}$ \\
		\hline
		$ \begin{array}{c}
		0\\
		1\\
		\\
		2\\
		\\
		\\
		3\\
		\\
		\\
		\\
\end{array}$
&$\begin{array}{c}
		0\\
		0\\
		1\\
		0\\
		1\\
		2\\
		0\\
		1\\
		2\\
		3\\
\end{array}$
&$\begin{array}{c}
0.03089963\\
0.09208239\\
0.09263068\\
0.15310157\\
0.15364899\\
0.15474383\\
0.21395707\\
0.21450363\\
0.21559675\\
0.21723640\\
\end{array}$
&$\begin{array}{c}
0.08051605\\
0.24044444\\
0.24099063\\
0.39912230\\
0.39966624\\
0.40075410\\
0.55655092\\
0.55709261\\
0.55817599\\
0.55980101\\	
\end{array}$
&$\begin{array}{c}
0.12410601\\
0.36996537\\
0.37050972\\
0.61277480\\
0.61331569\\
0.61439745\\
0.85254182\\
0.85307925\\
0.85415410\\
0.85576633\\
\end{array}$
& $\begin{array}{c}
0.14236930\\
0.42399536\\
0.42453895\\
0.70157568\\
0.70211529\\
0.70319450\\
0.97512297\\
0.97565862\\
0.97672990\\
0.97833679\\		
\end{array}$
&$\begin{array}{c}
0.14257464\\
0.42419921\\
0.42501458\\
0.70177804\\
0.70258745\\
0.70393644\\
0.97532384\\
0.97612730\\
0.97746640\\
0.97934108\\
\end{array}$
&$\begin{array}{c}
0.14291688\\
0.42453895\\
0.42562610\\
0.70211529\\
0.70319450\\
0.70481328\\
0.97565862\\
0.97672990\\
0.97833679\\
0.98047926\\
\end{array}$
&$\begin{array}{c}
0.14339599\\
0.42501458\\
0.42637351\\
0.70258745\\
0.70393644\\
0.70582499\\
0.97612730\\
0.97746640\\
0.97934108\\
0.98175131\\
\end{array}$
	\\
	\hline
	\hline
\end{tabular}
\end{table}	
\begin{table}[!h]
	\footnotesize
	\caption{${E_{nl}}$ (eV) of DFP for I${_2}$ molecule.}\label{tab:h2}
	\begin{tabular}{ccccccccc}
		\hline
		&&\multicolumn{3}{c}{${N=3}$} &${N=3}$ &\multicolumn{3}{c}{${\delta=1}$}
		\\
		\cmidrule(lr){3-5} \cmidrule(lr){7-9}
		${n}$ &	${l}$ & ${\delta=0.2}$ & ${\delta=0.5}$ & ${\delta=0.8}$ & ${\delta=1}$ & ${N=4}$ & ${N=5}$ & ${N=6}$ \\
		\hline
		$ \begin{array}{c}
		0\\
		1\\
		\\
		2\\
		\\
		\\
		3\\
		\\
		\\
		\\
		\end{array}$
		&$\begin{array}{c}
		0\\
		0\\
		1\\
		0\\
		1\\
		2\\
		0\\
		1\\
		2\\
		3\\
		\end{array}$
&$\begin{array}{c}
0.00290217\\
0.00866328\\
0.00868398\\
0.01441374\\
0.01443444\\
0.01447584\\
0.02015357\\
0.02017427\\
0.02021566\\
0.02027774\\	
\end{array}$
&$\begin{array}{c}
0.00757848\\
0.02266887\\
0.02268957\\
0.03768510\\
0.03770578\\
0.03774715\\
0.05262715\\
0.05264783\\
0.05268918\\
0.05275121\\
\end{array}$
&$\begin{array}{c}
0.01168805\\
0.03492991\\
0.03495060\\
0.05799444\\
0.05801512\\
0.05805647\\
0.08088166\\
0.08090232\\
0.08094365\\
0.08100563\\
\end{array}$
& $\begin{array}{c}
0.01341022\\
0.04005490\\
0.04007558\\
0.06646580\\
0.06648647\\.
0.06652781\\
0.09264296\\
0.09266361\\
0.09270493\\
0.09276689\\	
\end{array}$
&$\begin{array}{c}
0.01341798\\
0.04006266\\
0.04009368\\
0.06647355\\
0.06650456\\
0.06655623\\
0.09265070\\
0.09268169\\
0.09273333\\
0.09280562\\
\end{array}$
&$\begin{array}{c}
0.01343092\\
0.04007558\\
0.04011695\\
0.06648647\\
0.06652781\\
0.06658982\\
0.09266361\\
0.09270493\\
0.09276689\\
0.09284952\\
\end{array}$
&$\begin{array}{c}
0.01344903\\
0.04009368\\
0.04014539\\
0.06650456\\
0.06655623\\
0.06662857\\
0.09268169\\
0.09273333\\
0.09280562\\
0.09289857\\
\end{array}$
		\\
		\hline
		\hline
	\end{tabular}
\end{table}	
\begin{table}[!h]
	\footnotesize
	\caption{${E_{nl}}$ (eV) of DFP for NO molecule.}\label{tab:h3}
	\begin{tabular}{ccccccccc}
	\hline
	&&\multicolumn{3}{c}{${N=3}$} &${N=3}$ &\multicolumn{3}{c}{${\delta=1}$}
	\\
		\cmidrule(lr){3-5} \cmidrule(lr){7-9}
		${n}$ &	${l}$ & ${\delta=0.2}$ & ${\delta=0.5}$ & ${\delta=0.8}$ & ${\delta=1}$ & ${N=4}$ & ${N=5}$ & ${N=6}$ \\
		\hline
		$\begin{array}{c}
		0\\
		1\\
		\\
		2\\
		\\
		\\
		3\\
		\\
		\\
		\\
		\end{array}$
		&$\begin{array}{c}
		0\\
		0\\
		1\\
		0\\
		1\\
		2\\
		0\\
		1\\
		2\\
		3\\
		\end{array}$
&$\begin{array}{c}
0.02958454\\
0.08794744\\
0.08849444\\
0.14610481\\
0.14665108\\
0.14774361\\
0.20405658\\
0.20460211\\
0.20569316\\
0.20732971\\	
\end{array}$
&$\begin{array}{c}
0.07685720\\
0.22923250\\
0.22977770\\
0.38010256\\
0.38064583\\
0.38173235\\
0.52946844\\
0.53000977\\
0.53109243\\
0.53271639\\
\end{array}$
&$\begin{array}{c}
0.11839178\\
0.35239340\\
0.35293702\\
0.58275773\\
0.58329836\\
0.58437963\\
0.80949102\\
0.81002867\\
0.81110397\\
0.81271689\\
\end{array}$
& $\begin{array}{c}
0.13579494\\
0.40372140\\
0.40426436\\
0.66683706\\
0.66737660\\
0.66845566\\
0.92515259\\
0.92568870\\
0.92676092\\
0.92836923\\
\end{array}$
&$\begin{array}{c}
0.13599984\\
0.40392501\\
0.40473945\\
0.66703939\\
0.66784869\\
0.66919750\\
0.92535363\\
0.92615780\\
0.92749807\\
0.92937441\\
\end{array}$
&$\begin{array}{c}
0.13634133\\
0.40426436\\
0.40535028\\
0.66737660\\
0.66845566\\
0.67007422\\
0.92568870\\
0.92676092\\
0.92836923\\
0.93051361\\
\end{array}$
&$\begin{array}{c}
0.13681943\\
0.40473945\\
0.40609683\\
0.66784869\\
0.66919750\\
0.67108582\\
0.92615780\\
0.92749807\\
0.92937441\\
0.93178682\\
		\end{array}$
		\\
		\hline
		\hline
	\end{tabular}
\end{table}	
\begin{table}[!h]
	\footnotesize
	\caption{${E_{nl}}$ (eV) of DFP for N${_2}$ molecule.}\label{tab:h4}
	\begin{tabular}{ccccccccc}
		\hline
		&&\multicolumn{3}{c}{${N=3}$} &${N=3}$ &\multicolumn{3}{c}{${\delta=1}$}
		\\
		\cmidrule(lr){3-5} \cmidrule(lr){7-9}
		${n}$ &	${l}$ & ${\delta=0.2}$ & ${\delta=0.5}$ & ${\delta=0.8}$ & ${\delta=1}$ & ${N=4}$ & ${N=5}$ & ${N=6}$ \\
		\hline
		$ \begin{array}{c}
		0\\
		1\\
		\\
		2\\
		\\
		\\
		3\\
		\\
		\\
		\\
		\end{array}$
		&$\begin{array}{c}
		0\\
		0\\
		1\\
		0\\
		1\\
		2\\
		0\\
		1\\
		2\\
		3\\
		\end{array}$
&$\begin{array}{c}
0.03680223\\
0.10957917\\
0.11019288\\
0.18214266\\
0.18275555\\
0.18398134\\
0.25449260\\
0.25510469\\
0.25632884\\
0.25816506\\
\end{array}$
&$\begin{array}{c}
0.09579693\\
0.28598430\\
0.28659602\\
0.47458659\\
0.47519618\\
0.47641535\\
0.66160500\\
0.66221245\\
0.66342735\\
0.66524968\\
\end{array}$
&$\begin{array}{c}
0.14763020\\
0.43994427\\
0.44055425\\
0.72841609\\
0.72902278\\
0.73023614\\
1.01305263\\
1.01365603\\
1.01486282\\
1.01667299\\
\end{array}$
& $\begin{array}{c}
0.16934841\\
0.50415932\\
0.50476857\\
0.83388317\\
0.83448865\\
0.83569959\\
1.15853183\\
1.15913354\\
1.16033694\\
1.16214202\\
\end{array}$
&$\begin{array}{c}
0.16957830\\
0.50438779\\
0.50530166\\
0.83411023\\
0.83501843\\
0.83653210\\
1.15875747\\
1.15966002\\
1.16116427\\
1.16327018\\
		\end{array}$
&$\begin{array}{c}
0.16996144\\
0.50476857\\
0.50598707\\
0.83448865\\
0.83569959\\
0.83751597\\
1.15913354\\
1.16033694\\
1.16214202\\
1.16454875\\
\end{array}$
&$\begin{array}{c}
0.17049785\\
0.50530166\\
0.50682477\\
0.83501843\\
0.83653210\\
0.83865120\\
1.15966002\\
1.16116427\\
1.16327018\\
1.16597773\\
		\end{array}$
		\\
		\hline
		\hline
	\end{tabular}
\end{table}	
\begin{table}[!h]
	\footnotesize
	\caption{${-E_{nl}}$ (eV) of SDFP for CO molecule.}\label{tab:h6}
	\begin{tabular}{ccccccccc}
		\hline
		&&\multicolumn{3}{c}{${N=3}$}	 &${N=3}$&\multicolumn{3}{c}{${\delta=1}$}
		\\
		\cmidrule(lr){3-5} \cmidrule(lr){7-9}
		${n}$ &	${l}$ & ${\delta=0.2}$ & ${\delta=0.5}$ & ${\delta=0.8}$ & ${\delta=1}$ & ${N=4}$ & ${N=5}$ & ${N=6}$ \\
		\hline
		$ \begin{array}{c}
		0\\
		1\\
		\\
		2\\
		\\
		\\
		3\\
		\\
		\\
		\\
		\end{array}$
		&$\begin{array}{c}
		0\\
		0\\
		1\\
		0\\
		1\\
		2\\
		0\\
		1\\
		2\\
		3\\
		\end{array}$
		&$\begin{array}{c}
10.8141740\\
10.7529912\\
10.7524429\\
10.6919720\\
10.6914246\\
10.6903298\\
10.6311165\\
10.6305700\\
10.6294768\\
10.6278372\\
		\end{array}$
		&$\begin{array}{c}
10.7645575\\
10.6046292\\
10.6040830\\
10.4459513\\
10.4454074\\
10.4443195\\
10.2885227\\
10.2879810\\
10.2868976\\
10.2852726\\
		\end{array}$
		&$\begin{array}{c}
10.7209676\\
10.4751082\\
10.4745639\\
10.2322988\\
10.2317579\\
10.2306761\\
9.99253181\\
9.99199438\\
9.99091954\\
9.98930730\\
		\end{array}$
		& $\begin{array}{c}
10.7027043\\
10.4210782\\
10.4205347\\
10.1434979\\
10.1429583\\
10.1418791\\
9.86995067\\
9.86941502\\
9.86834374\\
9.86673685\\
		\end{array}$
		&$\begin{array}{c}
10.7024990\\
10.4208744\\
10.4200591\\
10.1432956\\
10.1424862\\
10.1411372\\
9.86974980\\
9.86894634\\
9.86760724\\
9.86573256\\
		\end{array}$
		&$\begin{array}{c}
10.7021568\\
10.4205347\\
10.4194475\\
10.1429583\\
10.1418791\\
10.1402604\\
9.86941502\\
9.86834374\\
9.86673685\\
9.86459438\\
		\end{array}$
		&$\begin{array}{c}
10.7016776\\
10.4200591\\
10.4187001\\
10.1424862\\
10.1411372\\
10.1392486\\
9.86894634\\
9.86760724\\
9.86573256\\
		9.86332233\\
		\end{array}$
		\\
		\hline
		\hline
	\end{tabular}
\end{table} 
\begin{table}[!h]
	\footnotesize
	\caption{${-E_{nl}}$ (eV) of SDFP for I${_2}$ molecule.}\label{tab:h7}
	\begin{tabular}{ccccccccc}
		\hline
		&&\multicolumn{3}{c}{${N=3}$}	 &${N=3}$&\multicolumn{3}{c}{${\delta=1}$}
		\\
		\cmidrule(lr){3-5} \cmidrule(lr){7-9}
		${n}$ &	${l}$ & ${\delta=0.2}$ & ${\delta=0.5}$ & ${\delta=0.8}$ & ${\delta=1}$ & ${N=4}$ & ${N=5}$ & ${N=6}$ \\
		\hline
		$ \begin{array}{c}
		0\\
		1\\
		\\
		2\\
		\\
		\\
		3\\
		\\
		\\
		\\
		\end{array}$
		&$\begin{array}{c}
		0\\
		0\\
		1\\
		0\\
		1\\
		2\\
		0\\
		1\\
		2\\
		3\\
		\end{array}$
		&$\begin{array}{c}
		1.55269782\\
		1.54693672\\
		1.54691602\\
		1.54118625\\
		1.54116556\\
		1.54112416\\
		1.53544642\\
		1.53542573\\
		1.53538434\\
		1.53532225\\
		\end{array}$
		&$\begin{array}{c}
		1.54802152\\
		1.53293112\\
		1.53291043\\
		1.51791490\\
		1.51789421\\
		1.51785284\\
		1.50297285\\
		1.50295217\\
		1.50291081\\
		1.50284878\\
		\end{array}$
		&$\begin{array}{c}
		1.54391195\\
		1.52067008\\
		1.52064939\\
		1.49760555\\
		1.49758488\\
		1.49754353\\
		1.47471833\\
		1.47469767\\
		1.47465635\\
		1.47459436\\
		\end{array}$
		& $\begin{array}{c}
		1.54218978\\
		1.51554509\\
		1.51552441\\
		1.48913420\\
		1.48911353\\
		1.48907219\\
		1.46295704\\
		1.46293638\\
		1.46289507\\
		1.46283310\\
		\end{array}$
		&$\begin{array}{c}
		1.54218201\\
		1.51553734\\
		1.51550631\\
		1.48912645\\
		1.48909544\\
		1.48904377\\
		1.46294929\\
		1.46291831\\
		1.46286667\\
		1.46279437\\
		\end{array}$
		&$\begin{array}{c}
	    1.54216908\\
	    1.51552441\\
	    1.51548304\\
	    1.48911353\\
	    1.48907219\\
	    1.48901018\\
	    1.46293638\\
	    1.46289507\\
	    1.46283310\\
	    1.46275048\\
		\end{array}$
		&$\begin{array}{c}
		1.54215097\\
		1.51550631\\
		1.51545460\\
		1.48909544\\
		1.48904377\\
		1.48897142\\
		1.46291831\\
		1.46286667\\
		1.46279437\\
		1.46270142\\
		\end{array}$
		\\
		\hline
		\hline
	\end{tabular}
\end{table} 
\begin{table}[!h]
	\footnotesize
	\caption{${-E_{nl}}$ (eV) of SDFP for NO molecule.}\label{tab:h8}
	\begin{tabular}{ccccccccc}
		\hline
		&&\multicolumn{3}{c}{${N=3}$}	 &${N=3}$&\multicolumn{3}{c}{${\delta=1}$}
		\\
		\cmidrule(lr){3-5} \cmidrule(lr){7-9}
		${n}$ &	${l}$ & ${\delta=0.2}$ & ${\delta=0.5}$ & ${\delta=0.8}$ & ${\delta=1}$ & ${N=4}$ & ${N=5}$ & ${N=6}$ \\
		\hline
		$ \begin{array}{c}
		0\\
		1\\
		\\
		2\\
		\\
		\\
		3\\
		\\
		\\
		\\
		\end{array}$
		&$\begin{array}{c}
		0\\
		0\\
		1\\
		0\\
		1\\
		2\\
		0\\
		1\\
		2\\
		3\\
		\end{array}$
		&$\begin{array}{c}
8.01414531\\
7.95578242\\
7.95523541\\
7.89762504\\
7.89707877\\
7.89598625\\
7.83967328\\
7.83912775\\
7.83803670\\
7.83640014\\
		\end{array}$
		&$\begin{array}{c}
7.96687265\\
7.81449736\\
7.81395215\\
7.66362729\\
7.66308403\\
7.66199750\\
7.51426141\\
7.51372008\\
7.51263743\\
7.51101347\\	
		\end{array}$
		&$\begin{array}{c}
7.92533807\\
7.69133645\\
7.69079283\\
7.46097213\\
7.46043149\\
7.45935023\\
7.23423884\\
7.23370118\\
7.23262589\\
7.23101296\\
		\end{array}$
		& $\begin{array}{c}
	7.90793492\\
	7.64000845\\
	7.63946549\\
	7.37689279\\
	7.37635326\\
	7.37527420\\
	7.11857727\\
	7.11804115\\
	7.11696893\\
	7.11536062\\
		\end{array}$
		&$\begin{array}{c}
	7.90773002\\
	7.63980484\\
	7.63899040\\
	7.37669047\\
	7.37588117\\
	7.37453235\\
	7.11837622\\
	7.11757206\\
	7.11623179\\
	7.11435544\\
		\end{array}$
		&$\begin{array}{c}
	7.90738852\\
	7.63946549\\
	7.63837958\\
	7.37635326\\
	7.37527420\\
	7.37365563\\
	7.11804115\\
	7.11696893\\
	7.11536062\\
	7.11321624\\
		\end{array}$
		&$\begin{array}{c}
	7.90691043\\
	7.63899040\\
	7.63763302\\
	7.37588117\\
	7.37453235\\
	7.37264404\\
	7.11757206\\
	7.11623179\\
	7.11435544\\
	7.11194304\\
		\end{array}$
		\\
		\hline
		\hline
	\end{tabular}
\end{table} 
\begin{table}[!h]
	\footnotesize
	\caption{${-E_{nl}}$ (eV) of SDFP for N${_2}$ molecule.}\label{tab:h9}
	\begin{tabular}{ccccccccc}
		\hline
		&&\multicolumn{3}{c}{${N=3}$}	 &${N=3}$&\multicolumn{3}{c}{${\delta=1}$}
		\\
		\cmidrule(lr){3-5} \cmidrule(lr){7-9}
		${n}$ &	${l}$ & ${\delta=0.2}$ & ${\delta=0.5}$ & ${\delta=0.8}$ & ${\delta=1}$ & ${N=4}$ & ${N=5}$ & ${N=6}$ \\
		\hline
		$ \begin{array}{c}
		0\\
		1\\
		\\
		2\\
		\\
		\\
		3\\
		\\
		\\
		\\
		\end{array}$
		&$\begin{array}{c}
		0\\
		0\\
		1\\
		0\\
		1\\
		2\\
		0\\
		1\\
		2\\
		3\\
		\end{array}$
		&$\begin{array}{c}
11.9013916\\
11.8286147\\
11.8280009\\
11.7560512\\
11.7554383\\
11.7542125\\
11.6837012\\
11.6830891\\
11.6818650\\
11.6800288\\	
		\end{array}$
		&$\begin{array}{c}
11.8423969\\
11.6522095\\
11.6515978\\
11.4636072\\
11.4629976\\
11.4617785\\
11.2765888\\
11.2759814\\
11.2747665\\
11.2729441\\
		\end{array}$
		&$\begin{array}{c}
11.7905636\\
11.4982495\\
11.4976396\\
11.2097777\\
11.2091710\\
11.2079577\\
10.9251412\\
10.9245378\\
10.9233310\\
10.9215208\\
		\end{array}$
		& $\begin{array}{c}
	11.7688454\\
	11.4340345\\
	11.4334252\\
	11.1043106\\
	11.1037052\\
	11.1024942\\
	10.7796620\\
	10.7790603\\
	10.7778569\\
	10.7760518\\
		\end{array}$
		&$\begin{array}{c}
	11.7686155\\
	11.4338060\\
	11.4328922\\
	11.1040836\\
	11.1031754\\
	11.1016617\\
	10.7794364\\
	10.7785338\\
	10.7770296\\
	10.7749236\\
		\end{array}$
		&$\begin{array}{c}
		11.7682324\\
		11.4334252\\
		11.4322067\\
		11.1037052\\
		11.1024942\\
		11.1006778\\
		10.7790603\\
		10.7778569\\
		10.7760518\\
		10.7736451\\
		\end{array}$
		&$\begin{array}{c}
		11.7676960\\
		11.4328922\\
		11.4313690\\
		11.1031754\\
		11.1016617\\
		11.0995426\\
		10.7785338\\
		10.7770296\\
		10.7749236\\
		10.7722161\\
		\end{array}$
		\\
		\hline
		\hline
	\end{tabular}
\end{table} 

\begin{figure}[!h]
	\begin{center}
		\begin{minipage}{.45\textwidth}
			(a) \centering
			\includegraphics[width=1.07\linewidth]{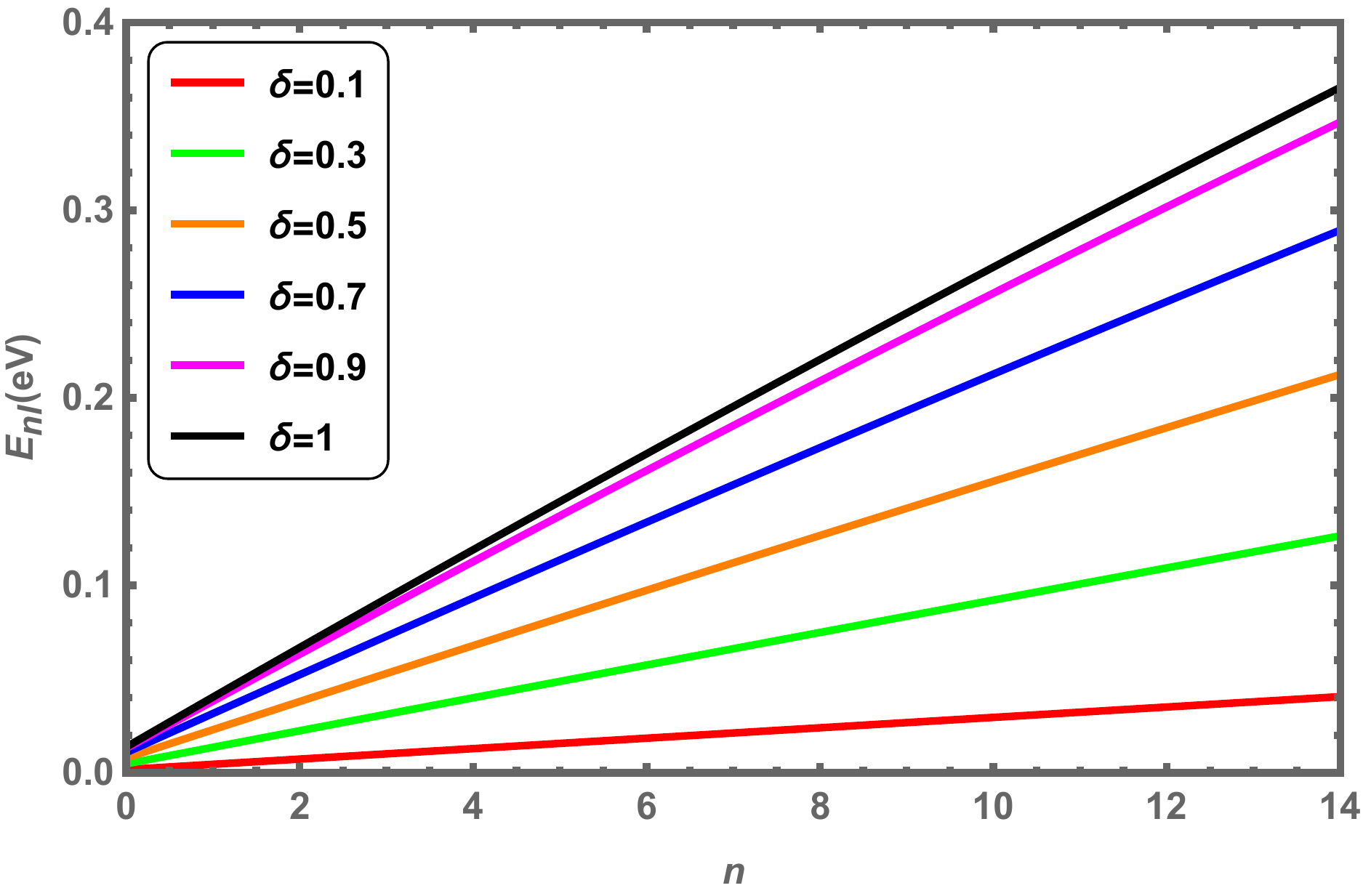}
		\end{minipage}
		\quad	
		\begin{minipage}{.45\textwidth}
			(b) \centering
			\includegraphics[width=1.08\linewidth]{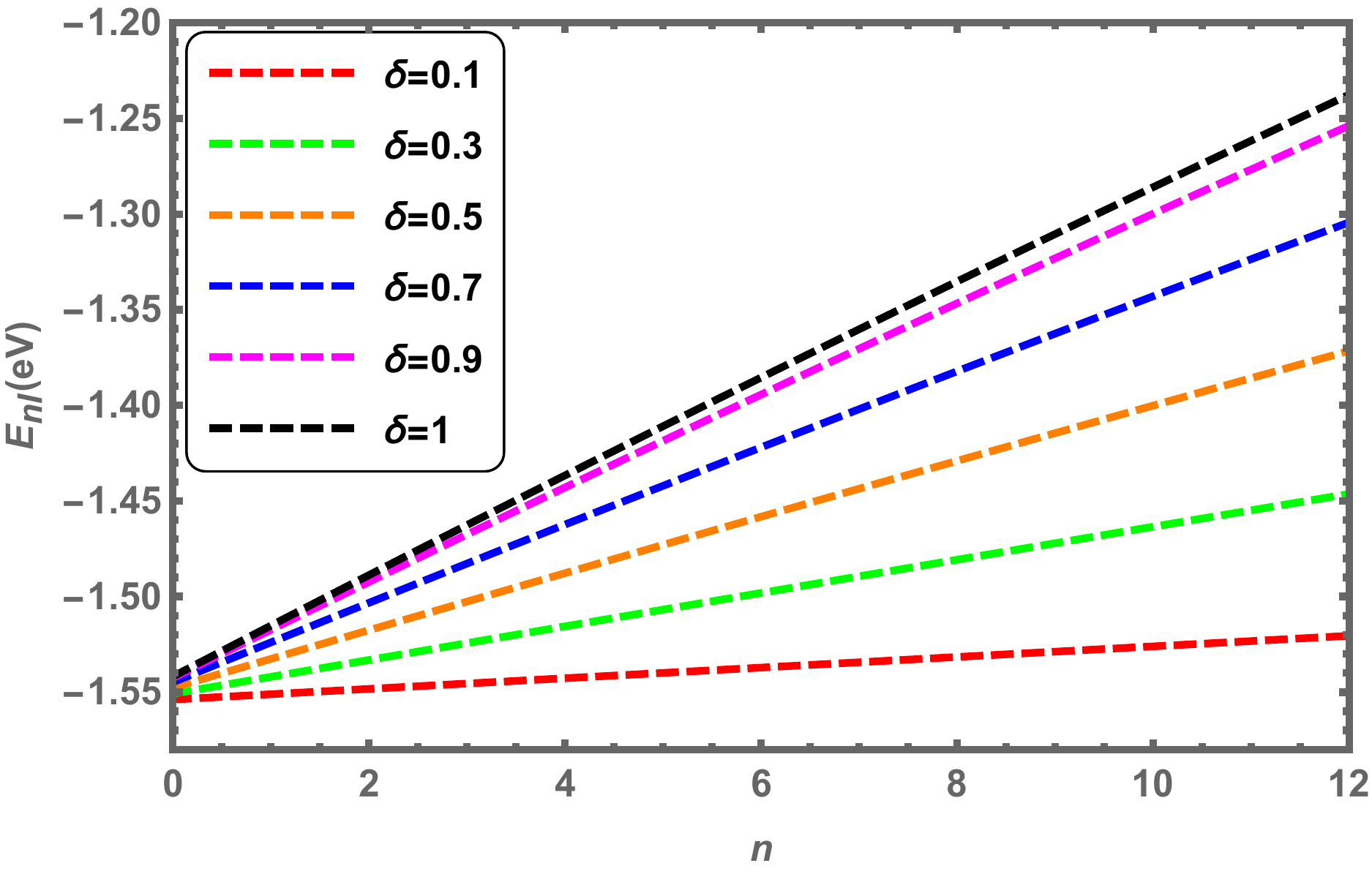}
		\end{minipage}
		\quad	
	\end{center}
	\caption{Energy levels of I${_2}$ molecule with ${n}$ at numerous values of ${\delta}$ for (a) DFP and (b) SDFP.}\label{f2}
\end{figure}
\begin{figure}[!h]
	\begin{center}
		\begin{minipage}{.45\textwidth}
			(a) \centering
			\includegraphics[width=1.07\linewidth]{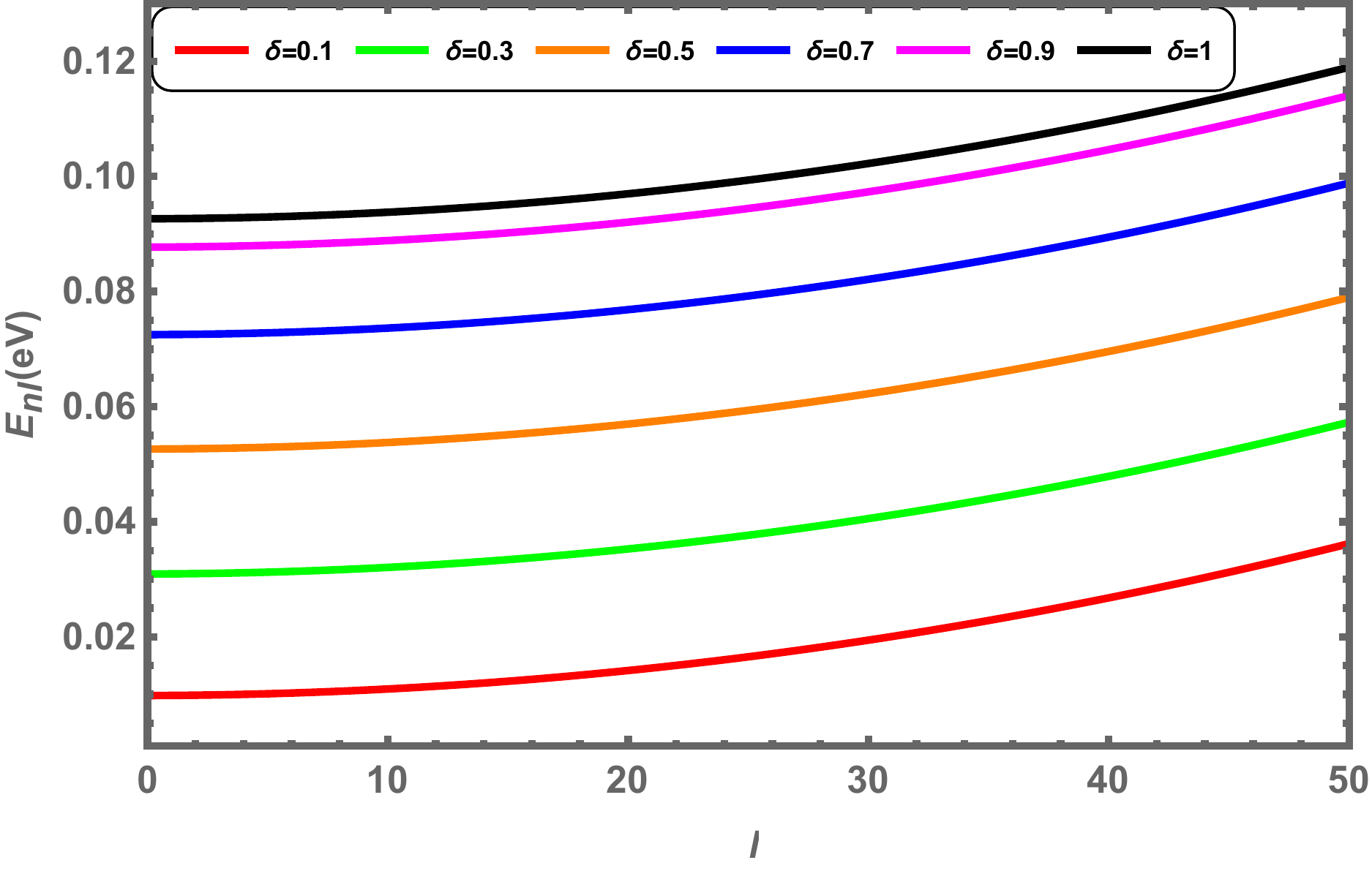}
		\end{minipage}
		\quad	
		\begin{minipage}{.45\textwidth}
			(b) \centering
			\includegraphics[width=1.08\linewidth]{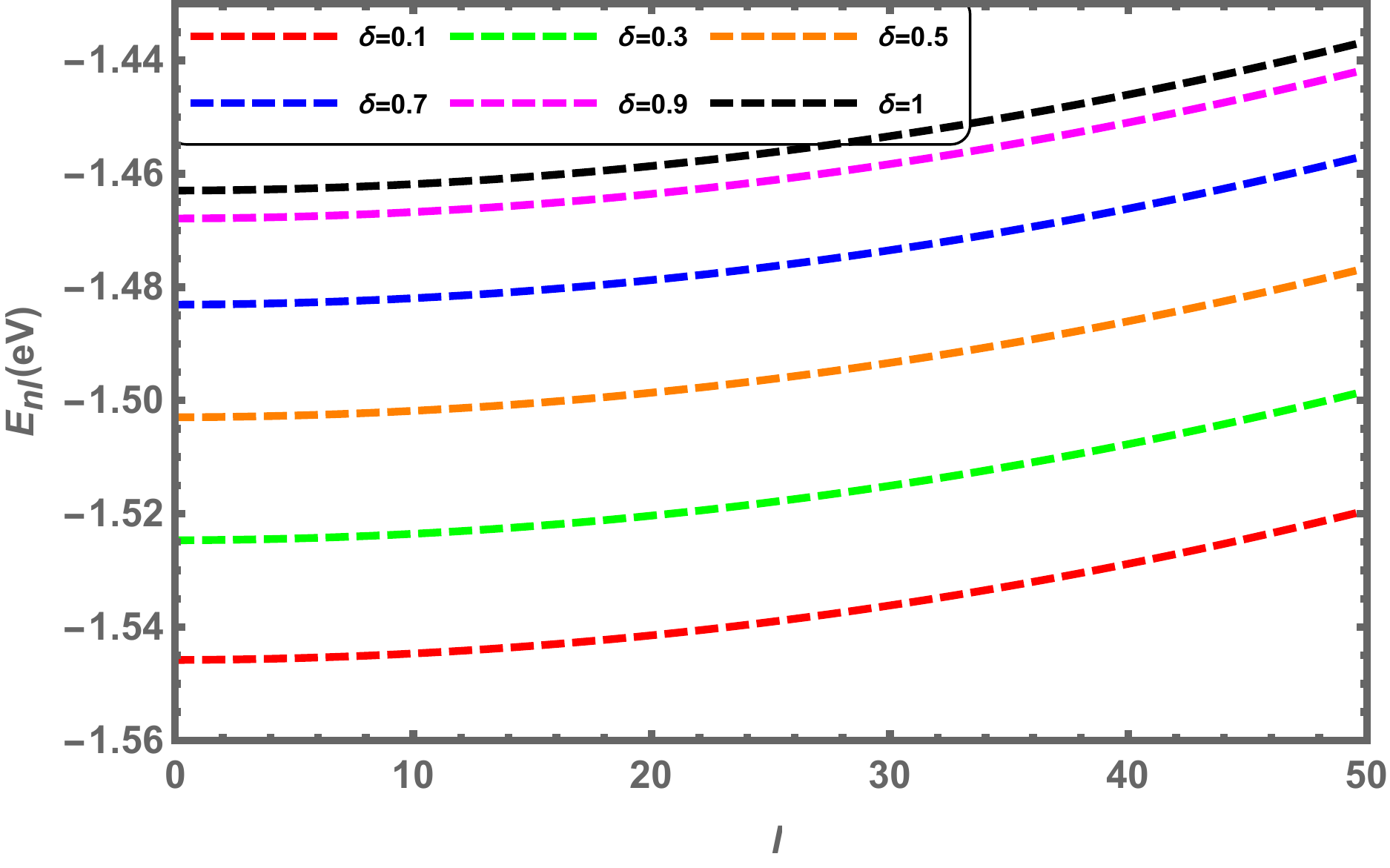}
		\end{minipage}
		\quad	
	\end{center}
	\caption{Energy levels of I${_2}$ molecule with ${l}$ at numerous values of ${\delta}$ for (a) DFP and (b) SDFP..}\label{f3}
\end{figure}
\begin{figure}[!h]
	\begin{center}
		\begin{minipage}{0.45\textwidth}
			(a) \centering
			\includegraphics[width=1.07\linewidth]{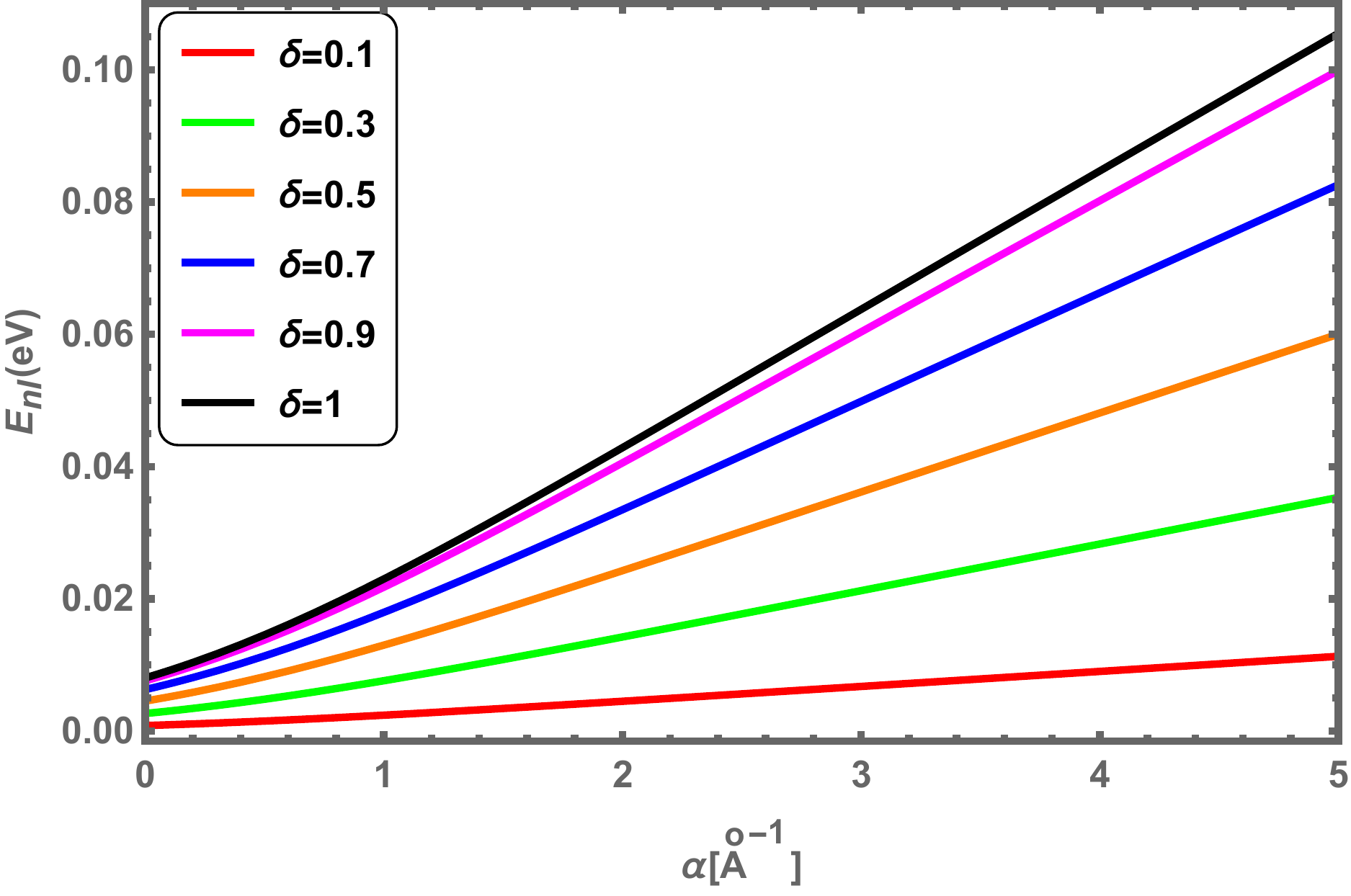}
		\end{minipage}
		\quad	
		\begin{minipage}{.45\textwidth}
			(b) \centering
			\includegraphics[width=1.08\linewidth]{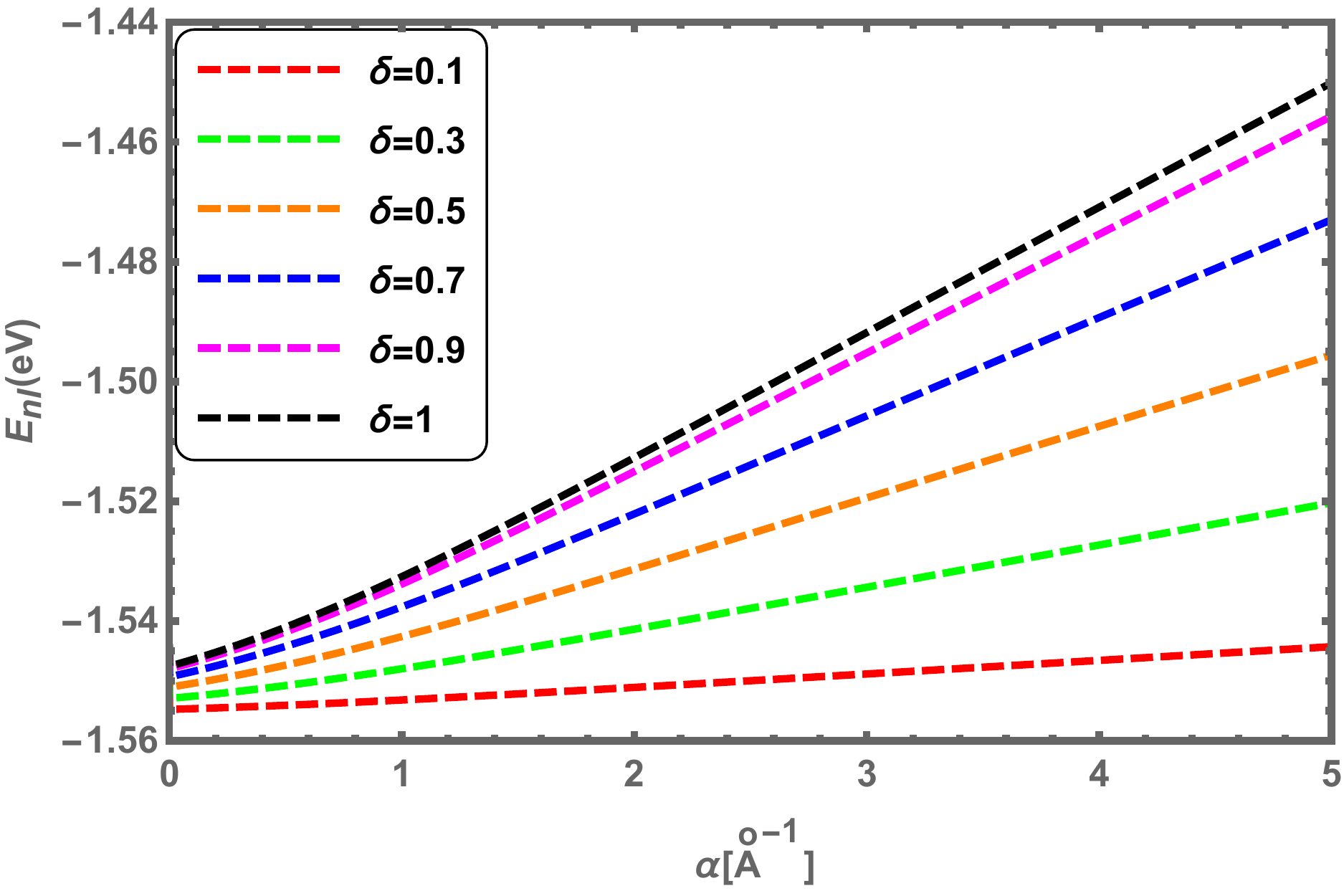}
		\end{minipage}
		\quad	
	\end{center}
	\caption{Energy levels of I${_2}$ molecule with ${\alpha}$ at numerous values of ${\delta}$ for (a) DFP and (b) SDFP.}\label{f4}
\end{figure}
\begin{figure}[!h]
	\begin{center}
		\begin{minipage}{.45\textwidth}
			(a) \centering
			\includegraphics[width=1.05\linewidth]{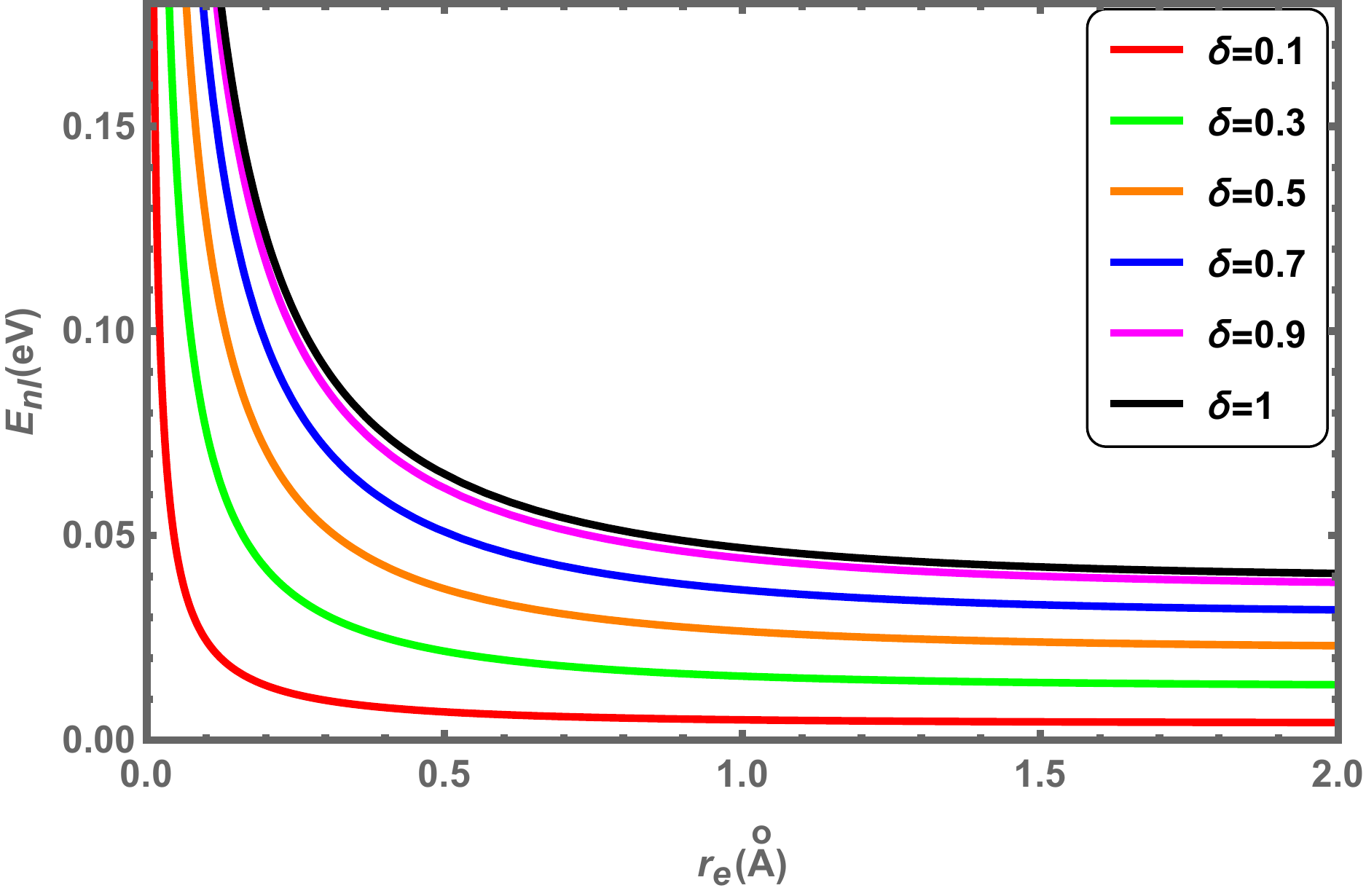}
		\end{minipage}
		\quad	
		\begin{minipage}{.45\textwidth}
			(b) \centering
			\includegraphics[width=1.08\linewidth]{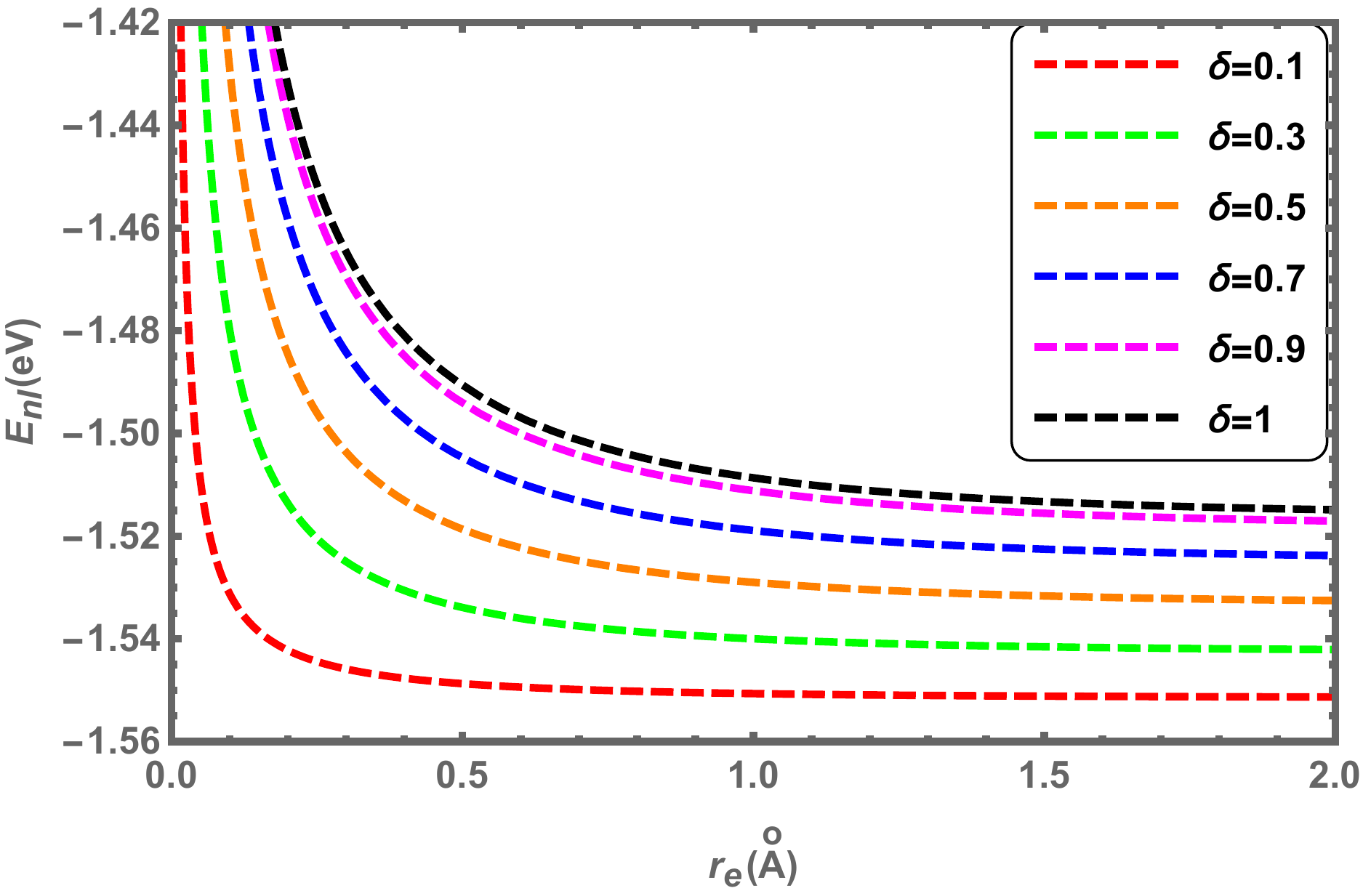}
		\end{minipage}
		\quad	
	\end{center}
	\caption{Energy levels of I${_2}$ molecule with ${r_e}$ at numerous values of ${\delta}$ for (a) DFP and (b) SDFP.}\label{f5}
\end{figure}
\begin{figure}[!h]
	\begin{center}
		\begin{minipage}{.45\textwidth}
			(a) \centering
			\includegraphics[width=1.05\linewidth]{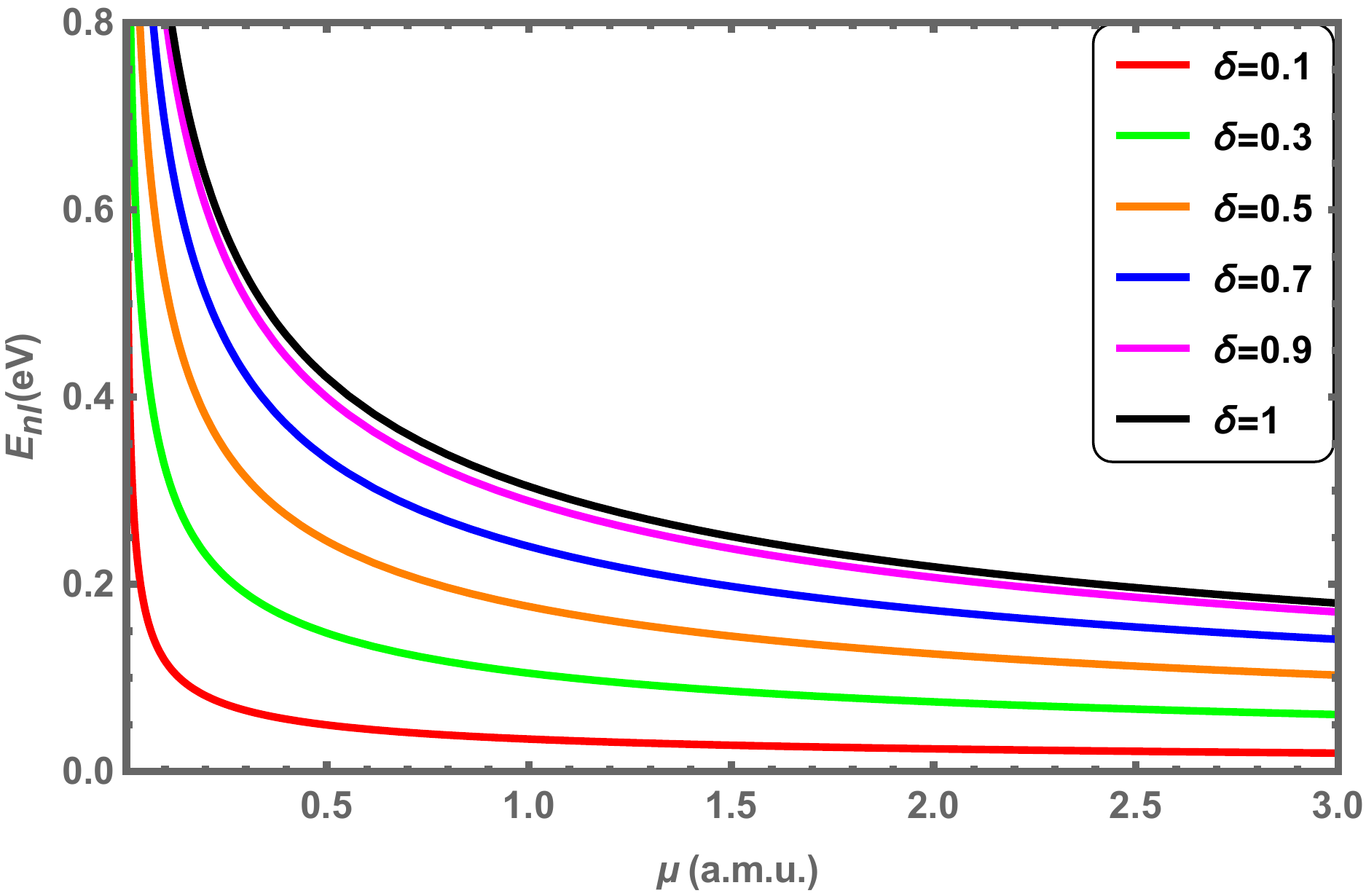}
		\end{minipage}
		\quad	
		\begin{minipage}{.45\textwidth}
			(b) \centering
			\includegraphics[width=1.08\linewidth]{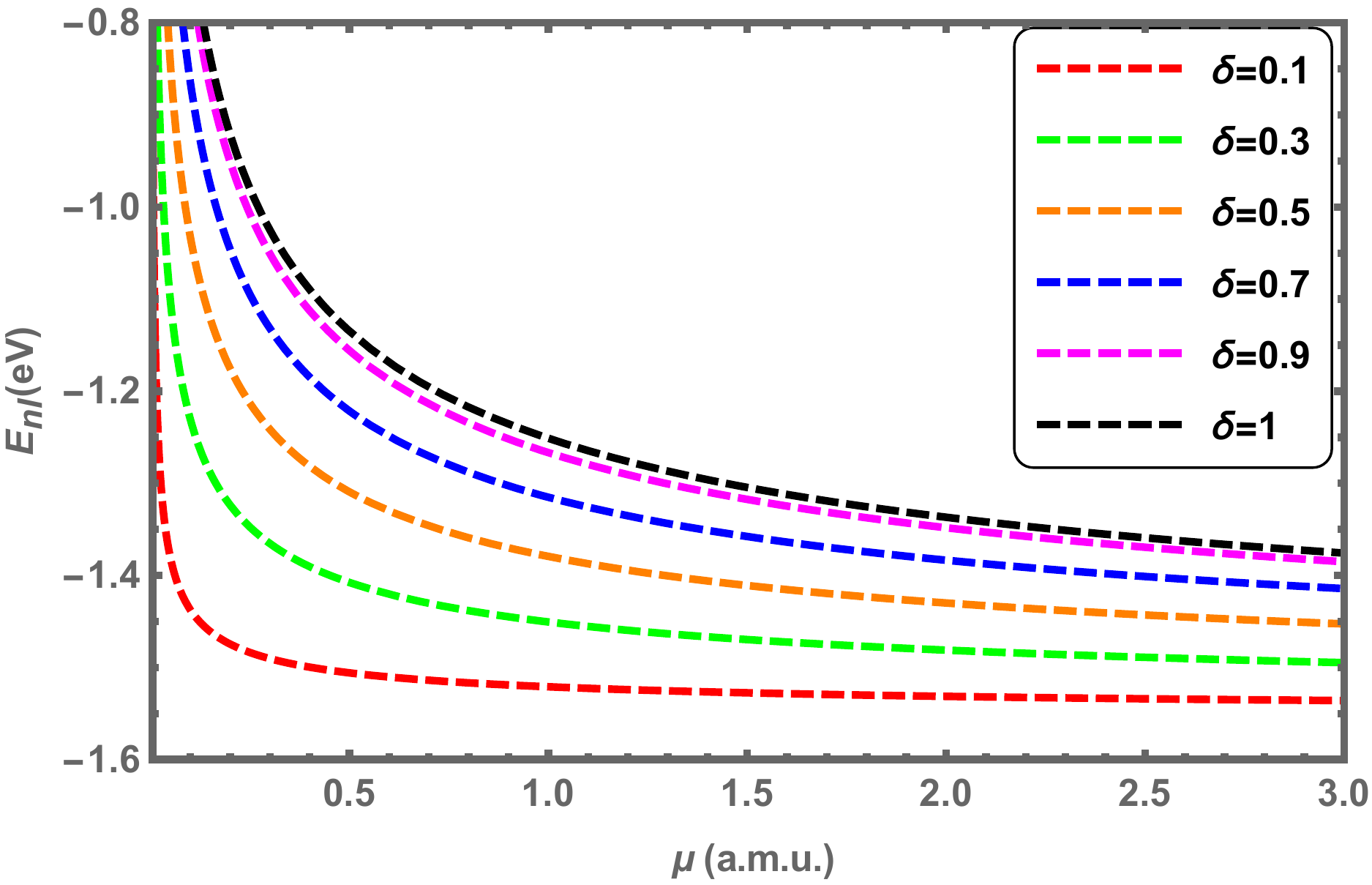}
		\end{minipage}
		\quad	
	\end{center}
	\caption{Energy levels of I${_2}$ molecule with ${\mu}$ at numerous values of ${\delta}$ for (a) DFP and (b) SDFP.}\label{f6}
\end{figure}
\begin{table}[!h]
	\footnotesize
	\centering
	\caption{${E_{nl}}$ (eV) of DFP for LiH and ScH molecules. }\label{tab:t2}
	\begin{tabular}{cccccccc}
		\hline
		&&\multicolumn{3}{c}{LiH}	 &\multicolumn{3}{c}{ScH}
		\\
		\cmidrule(lr){3-5} \cmidrule(lr){6-8}
		${n}$ &	${l}$	& Present  & \cite{NU3} & \cite{DF5} &  Present & \cite{NU3} & \cite{DF5}\\
		\hline
		$ \begin{array}{c}
		0\\
		1\\
		\\
		2\\
		\\
		\\
		3\\
		\\
		\\
		\\
		4\\
		\\
		\\
		\\
		\\
		5\\
		\\
		\\
		\\
		\\
		\\
		\end{array}$
		&$\begin{array}{c}
		0\\
		0\\
		1\\
		0\\
		1\\
		2\\
		0\\
		1\\
		2\\
		3\\
		0\\
		1\\
		2\\
		3\\
		4\\
		0\\
		1\\
		2\\
		3\\
		4\\
		5\\
		\end{array}$
		&$\begin{array}{c}
		0.103334307\\
		0.302004929\\
		0.303837626\\
		0.490684151\\
		0.492449049\\
		0.495977286\\
		0.669598629\\
		0.671297257\\
		0.674692995\\
		0.679782810\\
		0.838967494\\
		0.840601332\\
		0.843867529\\
		0.848763128\\
		0.855283702\\
		0.999002656\\
		1.000573134\\
		1.003712649\\
		1.008418320\\
		1.014685833\\
		1.022509444\\
		\end{array}$
		&$\begin{array}{c}
		0.103334650\\
		0.302005955\\
		0.303838653\\
		0.490685861\\
		0.492450759\\
		0.495978997\\
		0.669601019\\
		0.671299648\\
		0.674695388\\
		0.679785205\\
		0.838970564\\
		0.840604402\\
		0.843870601\\
		0.848766203\\
		0.855286782\\
		0.999006401\\
		1.000576880\\
		1.003716397\\
		1.008422072\\
		1.014689589\\
		1.022513206\\
		\end{array}$
		&$\begin{array}{c}
		0.103397595\\
		0.302068901\\
		0.303901531\\
		0.490748806\\
		0.492513638\\
		0.496041760\\
		0.669663964\\
		0.671362527\\
		0.674758151\\
		0.679847770\\
		0.839033509\\
		0.840667281\\
		0.843933363\\
		0.848828768\\
		0.855349091\\
		0.999069346\\
		1.000639759\\
		1.003779159\\
		1.008484637\\
		1.014751899\\
		1.022575209\\
		\end{array}$
		&$\begin{array}{c}
		0.104850693\\
		0.306246537\\
		0.307704128\\
		0.496950687\\
		0.498367396\\
		0.501200315\\
		0.677093198\\
		0.678469522\\
		0.681221677\\
		0.685348676\\
		0.846801672\\
		0.848138097\\
		0.850810460\\
		0.854817790\\
		0.860158632\\
		1.006201313\\
		1.007498315\\
		1.010091841\\
		1.013980933\\
		1.019164157\\
		1.025639600\\	
		\end{array}$
		&$\begin{array}{c}
		0.104850694\\
		0.306246538\\
		0.307704129\\
		0.496950687\\
		0.498367397\\
		0.501200316\\
		0.677093198\\
		0.678469522\\
		0.681221677\\
		0.685348677\\
		0.846801673\\
		0.848138097\\
		0.850810460\\
		0.854817791\\
		0.860158632\\
		1.006201313\\
		1.007498315\\
		1.010091841\\
		1.013980933\\
		1.019164157\\
		1.025639693\\
		\end{array}$
		&$\begin{array}{c}
		0.104938646\\
		0.306334489\\
		0.307791993\\
		0.497038637\\
		0.498455260\\
		0.501288002\\
		0.677181146\\
		0.678557384\\
		0.681309362\\
		0.685436100\\
		0.846889620\\
		0.848225957\\
		0.850898144\\
		0.854905213\\
		0.860245696\\
		1.006289259\\
		1.007586174\\
		1.010179523\\
		1.014068354\\
		1.019251220\\
		1.025726220\\
		\end{array}$
		\\
		\hline
		\hline
	\end{tabular}
\end{table}	
\begin{table}[!h]
	\footnotesize
	\centering
	\caption{ ${E_{nl}}$ (eV) of DFP for HCl and CO molecules. }\label{tab:t3}
	\begin{tabular}{ccccccc}
		\hline
		&&\multicolumn{3}{c}{HCl}	 &\multicolumn{2}{c}{CO}
		\\
		\cmidrule(lr){3-5} \cmidrule(lr){6-7}
		${n}$ &	${l}$	& Present & \cite{NU3} & \cite{DF16} & Present  & \cite{DF5}\\
		\hline
		$ \begin{array}{c}
		0\\
		1\\
		\\
		2\\
		\\
		\\
		3\\
		\\
		\\
		\\
		4\\
		\\
		\\
		\\
		\\
		5\\
		\\
		\\
		\\
		\\
		\\
		\end{array}$
		&$\begin{array}{c}
		0\\
		0\\
		1\\
		0\\
		1\\
		2\\
		0\\
		1\\
		2\\
		3\\
		0\\
		1\\
		2\\
		3\\
		4\\
		0\\
		1\\
		2\\
		3\\
		4\\
		5\\
		\end{array}$
&$\begin{array}{c}
0.201983505\\
0.590745819\\
0.593535604\\
0.960007699\\
0.962718245\\
0.968138299\\
1.310023186\\
1.312655523\\
1.317919173\\
1.325812091\\
1.641041232\\
1.643596366\\
1.648705629\\
1.656367005\\
1.666577476\\
1.953305816\\
1.955784737\\
1.960741588\\
1.968174386\\
1.978080159\\
1.990454948\\
\end{array}$
&$\begin{array}{c}
0.201984174\\
0.590747827\\
0.593537612\\
0.960011044\\
0.962721591\\
0.968141645\\
1.310027865\\
1.312660203\\
1.317923855\\
1.325816775\\
1.641047243\\
1.643602379\\
1.648711644\\
1.656373023\\
1.666583499\\
1.953313156\\
1.955792078\\
1.960748932\\
1.968181734\\
1.978087513\\
1.990462308\\
\end{array}$
&$\begin{array}{c}
0.202139134\\
0.590902787\\
0.593692401\\
0.960166004\\
0.962876380\\
0.968296136\\
1.310182826\\
1.312814994\\
1.318078347\\
1.325970803\\
1.641202203\\
1.643757170\\
1.648866136\\
1.656527052\\
1.666736911\\
1.953468116\\
1.955946871\\
1.960903424\\
1.968335762\\
1.978240925\\
1.990614950\\
\end{array}$
&$\begin{array}{c}
0.144850105\\
0.431437877\\
0.431981483\\
0.713979856\\
0.714519556\\
0.715598945\\
0.992488526\\
0.993024333\\
0.994095934\\
0.995703306\\
1.266976312\\
1.267508237\\
1.268572075\\
1.270167801\\
1.272295381\\
1.537455578\\
1.537983633\\
1.539039731\\
1.540623848\\
1.542735948\\
1.545375984\\
\end{array}$
&$\begin{array}{c}
0.144969072\\
0.430178995\\
0.430719631\\
0.711910800\\
0.712447561\\
0.713521071\\
0.989632352\\
0.990165249\\
0.991231033\\
0.992829678\\
1.263355977\\
1.263885023\\
1.264943104\\
1.266530195\\
1.268646261\\
1.533093944\\
1.533619151\\
1.534669552\\
1.536245125\\
1.538345834\\
1.540971631\\
\end{array}$
\\
		\hline
		\hline
	\end{tabular}
\end{table}	
\begin{table}[!h]
	\footnotesize
	\centering
	\caption{${E_{nl}}$ (eV) of DFP for HF, O$_{2}$ and H$_{2}$  molecules.}\label{tab:t4}
	\begin{tabular}{cccccccc}
		\hline
		&&\multicolumn{2}{c}{HF} &\multicolumn{2}{c}{O$_{2}$}
		&\multicolumn{2}{c}{H$_{2}$}
		\\
		\cmidrule(lr){3-4} \cmidrule(lr){5-6} \cmidrule(lr){7-8}
		${n}$ &	${l}$	& Present  & \cite{DF16} & Present & \cite{DF16} & Present & \cite{DF16}\\
		\hline
		$ \begin{array}{c}
		0\\
		1\\
		\\
		2\\
		\\
		\\
		3\\
		\\
		\\
		\\
		4\\
		\\
		\\
		\\
		\\
		5\\
		\\
		\\
		\\
		\\
		\\
		\end{array}$
		&$\begin{array}{c}
		0\\
		0\\
		1\\
		0\\
		1\\
		2\\
		0\\
		1\\
		2\\
		3\\
		0\\
		1\\
		2\\
		3\\
		4\\
		0\\
		1\\
		2\\
		3\\
		4\\
		5\\
		\end{array}$
			&$\begin{array}{c}
		0.291247434\\
		0.848733535\\
		0.853910381\\
		1.374650333\\
		1.379631880\\
		1.389590976\\
		1.869636743\\
		1.874426923\\
		1.884003381\\
		1.898358314\\
		2.334312319\\
		2.338914946\\
		2.348116388\\
		2.361909031\\
		2.380281458\\
		2.769277951\\
		2.773696723\\
		2.782530546\\
		2.795771983\\
		2.813409888\\
		2.835429422\\
		\end{array}$
		&$\begin{array}{c}
		0.296644754\\
		0.848791378\\
		0.853968904\\
		1.374741414\\
		1.379723603\\
		1.389683982\\
		1.869756985\\
		1.874547769\\
		1.884125435\\
		1.898482178\\
		2.334457762\\
		2.339060957\\
		2.348263535\\
		2.362057878\\
		2.380432567\\
		2.769444752\\
		2.773864056\\
		2.782698944\\
		2.795941975\\
		2.813582001\\
		2.835604179\\
		\end{array}$
		&$\begin{array}{c}
0.349980220\\
0.996777053\\
1.010323238\\
1.580248366\\
1.592700793\\
1.617539648\\
2.104086156\\
2.115507769\\
2.138289195\\
2.172307398\\
2.571680443\\
2.582129083\\
2.602968445\\
2.634083222\\
2.675301759\\
2.986148433\\
2.995677323\\
3.014680784\\
3.043050669\\
3.080625975\\
3.127194276\\	
\end{array}$
		&$\begin{array}{c}
0.365141571\\
0.998213655\\
1.011736491\\
1.582601950\\
1.595035333\\
1.619836697\\
2.107302268\\
2.118708590\\
2.141459973\\
2.175434452\\
2.575691305\\
2.586127919\\
2.606943719\\
2.638024369\\
2.679199648\\
2.990875171\\
3.000394872\\
3.019380399\\
3.047724487\\
3.085267435\\
3.131798515\\
\end{array}$		
&$\begin{array}{c}
0.101919228\\
0.302577426\\
0.303043144\\
0.499057714\\
0.499520138\\
0.500444979\\
0.691370318\\
0.691829458\\
0.692747730\\
0.694125117\\
0.879525427\\
0.879981291\\
0.880893009\\
0.882260567\\
0.884083941\\
1.063533190\\
1.063985785\\
1.064890966\\
1.066248719\\
1.068059018\\
1.070321833\\		
\end{array}$
		&$\begin{array}{c}
		0.102043850\\
		0.3015750557\\
		0.3020376575\\
		0.4974222630\\
		0.4978816054\\
		0.4988002821\\
		0.6891297288\\
		0.6895858196\\
		0.6904979933\\
		0.6918662339\\
		0.8767075389\\
		0.8771603861\\
		0.8780660724\\
		0.8794245822\\
		0.8812358915\\
		1.0601657396\\
		1.0606153510\\
		1.0615145658\\
		1.0628633684\\
		1.0646617349\\
		1.0669096339\\
		\end{array}$
			\\
		\hline
		\hline
	\end{tabular}
\end{table}	
\begin{table}[!h]
	\footnotesize
	\centering
	\caption{${-E_{nl}}$ (eV) of SDFP for CO and HCl molecules. }\label{tab:t5}
	\begin{tabular}{ccccccccc}
		\hline
	&&	\multicolumn{3}{c}{CO}	 &\multicolumn{3}{c}{HCl}
		\\
		\cmidrule(lr){3-5} \cmidrule(lr){6-8}
		${n}$ &	${l}$ & Present & \cite{DF1} & \cite{DF13} & Present & \cite{DF1} & \cite{DF13} \\
		\hline
$\begin{array}{c}
0\\
\\
\\
5\\
\\
\\
7\\
\\
\\
\end{array}$
&$\begin{array}{c}
0\\
5\\
10\\
0\\
5\\
10\\
0\\
5\\
10\\
\end{array}$
&$\begin{array}{c}
11.08074990\\
11.07253746\\
11.05064208\\
9.688144422\\
9.680224016\\
9.659107308\\
9.159162287\\
9.151357442\\
9.130548864\\
\end{array}$
		&$\begin{array}{c}
		11.08075178\\
		11.07253985\\
		11.05064581\\
		9.688146187\\
		9.680226284\\
		9.659110919\\
		9.159164003\\
		9.151359661\\
		9.130552425\\
\end{array}$
&$\begin{array}{c}
11.08074989\\
11.07360695\\
11.05456479\\
9.688144422\\
9.681355403\\
9.663257003\\
9.159162288\\
9.152513419\\
9.134788734\\
\end{array}$
&$\begin{array}{c}
4.417047400\\
4.374033861\\
4.259723835\\
2.665725089\\
2.628575957\\
2.529874313\\
2.096504515\\
2.061597436\\
1.968863331\\
\end{array}$
		&$\begin{array}{c}
		4.417077001\\
		4.374065784\\
		4.259761948\\
		2.665748019\\
		2.628601192\\
		2.529905688\\
		2.096524802\\
		2.061620020\\
		1.968892038\\
        \end{array}$
        &$\begin{array}{c}
       4.417047400\\
       4.378422991\\
       4.275898727\\
       2.665725089\\
       2.633922445\\
       2.549565348\\
       2.096504516\\
       2.067315382\\
       1.989918464\\
        \end{array}$
      \\   
		\hline
		\hline
	\end{tabular}
\end{table}	

\begin{table}[!h]
	\footnotesize
	\centering
	\caption{${-E_{nl}}$ (eV) of SDFP for LiH and ScH molecules. }\label{tab:t6}
	\begin{tabular}{cccccccc}
		\hline
		&&	\multicolumn{3}{c}{LiH}	 &\multicolumn{3}{c}{ScH}
		\\
		\cmidrule(lr){3-5} \cmidrule(lr){6-8}
		${n}$ &	${l}$	& Present  & \cite{DF1} & \cite{DF13} & Present & \cite{DF1} & \cite{DF3} \\
		\hline
		$\begin{array}{c}
		0\\
		\\
		\\
		5\\
		\\
		\\
		7\\
		\\
		\\
		\end{array}$
		&$\begin{array}{c}
		0\\
		5\\
		10\\
		0\\
		5\\
		10\\
		0\\
		5\\
		10\\
		\end{array}$
		&$\begin{array}{c}
	2.411932904\\
	2.383459166\\
	2.308127882\\
	1.516264555\\
	1.492757768\\
	1.430598166\\
	1.223382139\\
	1.201712023\\
	1.144423818\\
		\end{array}$
		&$\begin{array}{c}
		2.411949045\\
		2.383476249\\
		2.308147473\\
		1.516277294\\
		1.492771433\\
		1.430614300\\
		1.223393538\\
		1.201724343\\
		1.144438594\\
		\end{array}$
		&$\begin{array}{c}
		2.411932904\\
		2.384580437\\
		2.312283846\\
		1.516264555\\
		1.494210023\\
		1.435969432\\
		1.223382139\\
		1.203290365\\
		1.150257886\\
	    \end{array}$
		&$\begin{array}{c}
	2.145149306\\
	2.122682696\\
	2.062960535\\
	1.243798687\\
	1.224360400\\
	1.172700118\\
	0.955436772\\
	0.937159725\\
	0.888590944\\
		\end{array}$
		&$\begin{array}{c}
		2.1451493060\\
		2.1226840150\\
		2.0629653710\\
		1.2437986870\\
		1.2243617190\\
		1.1727049540\\
		0.9554367725\\
		0.9371610441\\
		0.8885957808\\
		\end{array}$
		&$\begin{array}{c}
	2.145149306\\
	2.122682696\\
	-----\\
	-----\\
	-----\\
	-----\\
	-----\\
	-----\\
	-----\\
		\end{array}$
				\\
		\hline
		\hline
	\end{tabular}
\end{table}	
\begin{table}[!h]
	\footnotesize
	\centering
	\caption{${-E_{nl}}$ (eV) of SDFP for H$_{2}$ and I$_{2}$ molecules.}\label{tab:t7}
	\begin{tabular}{cccccccc}
		\hline
		&&	\multicolumn{3}{c}{H$_{2}$}	 &\multicolumn{3}{c}{I$_{2}$}
		\\
		\cmidrule(lr){3-5} \cmidrule(lr){6-8}
		${n}$ &	${l}$	& Present  & \cite{DF1} & \cite{DF13} & Present & \cite{DF1} & \cite{DF3}\\
		\hline
		$\begin{array}{c}
		0\\
		\\
		\\
		5\\
		\\
		\\
		7\\
		\\
		\\
		\end{array}$
		&$\begin{array}{c}
		0\\
		5\\
		10\\
		0\\
		5\\
		10\\
		0\\
		5\\
		10\\
		\end{array}$
		&$\begin{array}{c}
	4.394619779\\
	4.176613157\\
	3.621820491\\
	1.758451567\\
	1.617405724\\
	1.260433707\\
	1.077636993\\
	0.961809891\\
	0.669826131\\
		\end{array}$
		&$\begin{array}{c}
		4.394619779\\
		4.176618048\\
		3.621838424\\
		1.758451567\\
		1.617410615\\
		1.260451640\\
		1.077636993\\
		0.961814782\\
		0.669844065\\
		\end{array}$
			&$\begin{array}{c}
	4.394619776\\
	4.180734836\\
	3.637874575\\
	1.758451566\\
	1.624065481\\
	1.285780199\\
	1.077636991\\
	0.969361194\\
	0.698426418\\
		    \end{array}$
		&$\begin{array}{c}
	1.542189775\\
	1.541879304\\
	1.541051381\\
	1.411303767\\
	1.410994355\\
	1.410169257\\
	1.360584943\\
	1.360275954\\
	1.359451985\\
		\end{array}$
		&$\begin{array}{c}
		1.542189775\\
		1.541879340\\
		1.541051513\\
		1.411303767\\
		1.410994391\\
		1.410169388\\
		1.360584943\\
		1.360275990\\
		1.359452116\\
		\end{array}$
		&$\begin{array}{c}
		1.542189775\\
		1.541879304\\
		-----\\
		-----\\
		-----\\
		-----\\
		-----\\
		-----\\
		-----\\
		\end{array}$
		\\
		\hline
		\hline
	\end{tabular}
\end{table}	
\section{Results and discussion }
In this part, we employ the previously obtained solutions to analyze the energy spectra of numerous DMs for both the DFP and the SDFP under the influence of the fractional parameter (${\delta}$) in ${N}$-dimensional space. In our computations, we utilize the spectroscopic values reported in Table (\ref{tab:t0}) for selected DMs: HCl, CO, O${_2}$ and NO besides the following replacements \cite{Convs.}: ${\hbar c=1973.29}$ eV ${\overset{o}{A}}$,
${1cm^{-1}=1.239841875\times10^{-4} }$ eV, and ${1amu=931.494028}$ MeV/ ${c^2}$. In Fig. (\ref{f1}), the variation of the DFP and its shifted is indicated at ${r=r_e}$ for some DMs.
In Tables (\ref{tab:h1}- \ref{tab:h4}), the numerical results of the energy spectra for various DMs: CO, I${_2}$, NO and N${_2}$ with the DFP are displayed at diverse values of the fractional parameter (${\delta}$) and the dimensional number (${N}$). It is found that increasing the fractional parameter (${\delta}$) causes a significant increase in the energy eigenvalues. Concurrently, as the dimensional number (${N}$) rises, so do the energy eigenvalues. Furthermore, the impact of fractional parameter (${\delta}$) and the dimensional number (${N}$) on the energy eigenvalues for the SDFP is explored in Tables (\ref{tab:h6}- \ref{tab:h9}). We also observe here that the energy eigenvalues move to higher values as ${\delta}$ and ${N}$ rise. The influence of the fractional parameter on the behavior of the energy spectra for both the DFP and SDFP with varied potential parameters is depicted in Figs. (\ref{f2}- \ref{f6}) for a clearer appreciation. For various values of the fractional parameter ${\delta=0.1, \delta=0.3,\delta=0.5,\delta=0.7, \delta=0.9}$ and ${\delta=1}$, the variations of the energy spectra with the vibrational (${n}$) and rotational (${l}$) quantum numbers are shown in Fig. (\ref{f2}) and Fig. (\ref{f3}) respectively. Fig. (\ref{f2}) reveals that the energy spectra gradually grow as the vibrational quantum number (${n}$) increases. In line with this inspection, Fig. 3 illustrates a progressive rise also in the energy spectra as the rotational quantum number (${l}$) rises. In Fig. (\ref{f4}), we display the behavior of the energy spectra with the screening parameter (${\alpha}$) for different values of the fractional parameter. As seen in Fig. (\ref{f4}), the energy spectra elevate in a consistent manner as the screening parameter is increased. In Fig. (\ref{f5}), the energy spectra are plotted with the equilibrium bond length (${r_e}$) at ${\delta=0.1, \delta=0.3,\delta=0.5,\delta=0.7, \delta=0.9}$ and ${\delta=1}$. As shown in Fig. (\ref{f5}), the energy spectra decrease by increasing the equilibrium bond length. The behavior of the energy spectra with the reduced mass (${\mu}$) is depicted in Fig. (\ref{f6}). Obviously, increasing the reduced mass leads to a drop in the energy spectra. 
\\
\\
\\
\\
\\
Due to our investigation being innovative, we are unable to compare the obtained results for different values of the fractional parameter. As a consequence, we recover the classical solutions of the DFP and its shifted by setting ${\delta=1}$ and comparing our results to those found in the literature. In Tables (\ref{tab:t2}- \ref{tab:t4}), we report the numerical results for the ro-vibrational energy spectra of the DFP for  several DMs: LiH, ScH, HCl, CO, O${_2}$ and H${_2}$ in comparison with the found in Refs. \cite{DF16}, \cite{NU3}, \cite{DF5}. 
Furthermore, the ro-vibrational energy spectra of the SDFP for numerous DMs: LiH, ScH, HCl, CO, H${_2}$ and I${_2}$ are displayed in Tables (\ref{tab:t5}- \ref{tab:t7}) compared to the findings in Refs. \cite{DF1}, \cite{DF3}, \cite{DF13}.
As can be shown in Tables (\ref{tab:t2}- \ref{tab:t7}), the ro-vibrational energy spectra of all selected DMs rise as the vibrational and rotational quantum numbers increase. Importantly, one can see that our estimates are perfectly consistent with prior works that used other techniques. 
To our knowledge, such an investigation has never been done previously. We hope that the current findings will be helpful in future studies.

 \newpage
\section{Conclusion}
In this paper, we have employed the generalized fractional derivative (GFD) to derive the bound state solutions of the ${N}$-dimensional SE with the DFP within the framework of the generalized fractional NU method. The formulas of energy eigenvalues and corresponding eigenfunctions for the DFP as a function of the fractional parameter have been produced for arbitrary values of the vibrational and rotational quantum numbers in the ${N}$-dimensional space. As an application, we have evaluated the ro-vibrational energy levels of varied diatomic molecules (DMs) for both the DFP and its shifted potential at different values of the fractional parameter. The effect of the fractional parameter on the energy levels has also been graphically illustrated. We noted that decreasing fractional parameter (${\delta}$) lowers the energy eigenvalues significantly. As a consequence, we conclude that the diatomic molecule's energy is more bounded at lower fractional parameter values than in the classical case (${\delta=1}$). Furthermore, we estimated the energy levels of numerous DMs in three-dimensional space as well as higher dimensions. It's notable that as the number of dimensions grows, the energy spectrum grows as well. The variation of the energy levels of the DFP and its shifted potential for the I${_2}$ DM has been graphically depicted with the reduced mass, screening parameter, equilibrium bond length, rotational and vibrational quantum numbers. To prove the validity of our findings, we calculated the energy levels of numerous DMs at ${\delta=1}$ and compared the results to previous studies. As can be observed, our findings are in perfect accord with those of others. We infer that the fractional parameter has a significant impact on the ro-vibrational energy levels of DMs. Consequently, exploring the solutions of the ${N}$-dimensional SE within the framework of the GFD, and studying the impact of the fractional number on various features of diverse systems in molecular and atomic physics should be investigated. Finally, we indicate that this work would be extended in the future to derive the solutions of the ${N}$-dimensional SE within the scope of the GFD for further molecular potentials such as the Kratzer potential, Morse potential, Pöschl-Teller potential, Hulthén potential, and so on.

\end{document}